\documentclass[12pt]{article}

\usepackage{setspace}
\usepackage{scicite}
\usepackage{amssymb}

\usepackage{graphics,graphicx}

\usepackage{times}

\newenvironment{sciabstract}{%
\begin{quote} \bf}
{\end{quote}}

\newcounter{lastnote}
\newenvironment{scilastnote}{%
\setcounter{lastnote}{\value{enumiv}}%
\addtocounter{lastnote}{+1}%
\begin{list}%
{\arabic{lastnote}.}
{\setlength{\leftmargin}{.22in}}
{\setlength{\labelsep}{.5em}}}
{\end{list}}

\def\xmm{{\it XMM-Newton~\/}}

\def\suzaku{{\it Suzaku}}

\def\swift{{\it SWIFT~\/}}

\def\epicpn{{\it EPIC}{\rm-pn}}
\def\epicmos1{{\it EPIC}{\rm-MOS1~\/}}
\def\epicmos2{{\it EPIC}{\rm-MOS2 ~\/}}
\def\epicmos{{\it EPIC}{\rm-MOS}}

\def\xmm{{\it XMM-Newton}}
\def\swift{{\it SWIFT}}

\def\xspec{\hbox{\sc XSPEC}}

\def\heasoftv{\hbox{\rm HEASOFT\thinspace v6.6.1}}
\def\xselect{\hbox{\rm XSELECT}}
\def\ftool{\hbox{\rm FTOOL}}

\def\s{\hbox{$\rm\thinspace s$}}
\def\ks{\hbox{$\rm\thinspace ks$}}

\def\pcmsq{\hbox{$\rm\thinspace cm^{-2}$}}

\def\kev{\hbox{$\rm\thinspace keV$}}

\def\ctsps{\hbox{$\rm\thinspace count~s^{-1}$}}

\def\ergpcmsqps{\hbox{$\rm\thinspace erg~cm^{-2}~s^{-1}$}}

\def\ergps{\hbox{$\rm\thinspace erg~s^{-1}$}}

\def\msun{\hbox{$\rm\thinspace M_{\odot}$}}

\def\nh{${\it N}_{\rm H}$}

\def\chisq{{\chi^{2}}}

\def\zpo{\rm{\small ZPOWERLAW}}

\def\zphabs{\rm{\small ZPHABS}}
\def\diskbb{\rm{\small DISKBB}}

\def\xspec{\hbox{\small XSPEC~\/}}

\def\heasoftv{\hbox{\rm{\small HEASOFT}~v6.11.1\/}}
\def\xselect{\hbox{\rm{\small XSELECT~\/}}}

\def\ftool{\hbox{\rm{\small FTOOL}}}

\def\rmfgen{\hbox{\rm{\small RMFGEN}}}
\def\arfgen{\hbox{\rm{\small ARFGEN}}}

\def\grid25{\hbox{\rm{\small GRID25}}}
\def\aeattcor{\hbox{\rm{\small AEATTCOR}}}
\def\pile_est{\hbox{\rm{\small PILE_EST}}}
\def\xistime{\hbox{\rm{\small XISTIME}}}
\def\xiscoord{\hbox{\rm{\small XISCOORD}}}
\def\pharbn{\hbox{\rm{\small PHARBN}}}
\def\epatplot{\hbox{\rm{\small EPATPLOT}}}

\def\msun{\hbox{$\rm\thinspace M_{\odot}$}}

\def\suzaku{{\it Suzaku}}
\def\xmm{{\it XMM-Newton}}

\def\s{\hbox{$\rm\thinspace s$}}
\def\ks{\hbox{$\rm\thinspace ks$}}

\def\pcmsq{\hbox{$\rm\thinspace cm^{-2}$}}

\def\kev{\hbox{$\rm\thinspace keV$}}

\def\ctsps{\hbox{$\rm\thinspace count~s^{-1}$}}

\def\ergpcmsqps{\hbox{$\rm\thinspace erg~cm^{-2}~s^{-1}$}}

\def\ergps{\hbox{$\rm\thinspace erg~s^{-1}$}}

\def\msun{\hbox{$\rm\thinspace M_{\odot}$}}

\def\j{\hbox{\rm Sw J1644+57}}
\def\x{{\it XMM~{\rm \#1}}}

\title{A  200-s Quasi-Periodicity Following the Tidal Disruption of a Star by a Dormant Black Hole}

\author
{R.~C.~Reis,$^{1\ast}$  J.~M.~Miller,$^{1}$, M.~T.~Reynolds,$^{1}$,  K.~G\"{u}ltekin,$^{1}$ , D.~Maitra,$^{1}$ , A.~L.~King,$^{1}$  \\ T. E. Strohmayer$^{2}$\\
\\
\normalsize{$^{1}$Department of Astronomy, University of Michigan,}\\
\normalsize{Ann Arbor, Michigan 48109, USA }\\
\normalsize{$^{2}$Astrophysics Science Division, NASA Goddard Space Flight Center,}\\
\normalsize{Greenbelt, MD 20771, USA }\\
\\
\normalsize{$^\ast$To whom correspondences should be addressed. E-mail:  rdosreis@umich.edu.}
}

\date{}

\begin{document} 


\baselineskip24pt


\maketitle

\label{abstract}

\begin{sciabstract}
Supermassive black holes (SMBHs; $M\gtrsim10^5\msun$) are known to exist at the centre of most  galaxies with sufficient stellar mass. In the local Universe, it is possible to infer their properties from the surrounding stars or gas. However, at high redshifts we require active, continuous accretion to infer the presence of the SMBHs, often coming in the form of long-term accretion in active galactic nuclei. SMBHs can also capture and tidally disrupt stars orbiting nearby, resulting in bright flares from otherwise quiescent black holes. Here, we report on a $\sim200$-s X-ray quasi-periodicity around a previously dormant SMBH located in the centre of a galaxy at redshift $z=0.3534$. This result may open the  possibility of probing general relativity beyond our local Universe.
\end{sciabstract}

\label{intro}

Tidal disruption of stars as a means to fuel active SMBHs was originally proposed in 1975\cite{hills1975} but it was not until over a decade later that the possibility of using the expected electromagnetic flares to study non-active SMBHs was proposed\cite{rees1988}.  Since then, various candidate tidal disruption flares (TDFs)  have been identified based on luminous flares observed from optical to X-rays\cite{Komossa1999,GreineSchwarzr2000,Gezari2006,Gezari2008,vanVelzen2011}, but it is only in recent years that such objects have been confirmed through the observation of relativistic flares\cite{Burrows2011Nature, Levan2011Sci,Bloom2011Sci, Zauderer2011Natur, Cenko2012}.


Swift J164449.3+573451 (hereafter \j)  was detected\cite{Burrows2011Nature} by the \textit{Swift} Burst Alert Telescope (BAT, 15-150\kev) on 28 March 2011 as it reached  X-ray luminosities greater than $\sim10^{48}\ergps$. Prompt, multi-wavelength observations spanning from radio to  $\gamma-$rays\cite{Zauderer2011Natur,Levan2011Sci} confirmed the position of the source to be coincidental with the nucleus of an inactive galaxy and showed the presence of rapid time variability ($\approx10^2\s$) at high energies\cite{Burrows2011Nature,Bloom2011Sci} , indicative of a highly compact source of emission ($\sim 30~{\rm million~km}$ or $ \sim 0.2~{\rm AU}$). Furthermore, the  detection of a relativistic and highly collimated radio jet, for the first time associated with a TDF\cite{Bloom2011Sci,Zauderer2011Natur}, makes \j\ analogous to a small-scale blazar\cite{Bloom2011Sci}.   Following the identification of this source as a tidal disruption candidate (TDC), we initiated a series of  twelve, bi-weekly observations with the large X-ray satellite \xmm, starting approximately 19 days after the BAT trigger  (Fig. 1), together with a further \suzaku\ pointing $\sim10$ days earlier\cite{supplementary}. The long-term evolution in the X-ray (0.2-10\kev; observed frame) luminosity followed roughly the expected rate of mass return to the black hole [$\propto t^{-5/3}$ for TDFs\cite{rees1988}] and the short-term, rapid variability was also  apparent.

\label{powerspec}
We produced light curves for all observations of \j\ over the 0.2-10.0\kev\ energy band and Fourier transformed these in order to obtain various power density spectra.  A least-squares fits to the soft energy band ($<2\kev$) power spectrum for the \suzaku\ observation is consistent with a band-limited (red) noise component, described by a  $\Gamma\approx-1.8$ power-law  plus Poisson (white) noise (Fig. S1 and Table~S1). However, the 2-10\kev\ power spectra of both the \suzaku\ and first \xmm\ observations (hereafter \x) displayed  a potential Quasi-Periodic Oscillation (QPO) component near 5mHz (Fig. 2). This feature  has a centroid frequency of $\nu\sim 4.8~$mHz and frequency width (full-width at half-maximum) of $\delta\nu_{\rm suzaku} \leq0.4$~mHz  and $\delta\nu_{\rm xmm}=0.3$~mHz (quality factor $Q=\nu/\delta\nu \geq 12$ and $\sim 15$ for \suzaku\ and \x\  respectively). The fractional root-mean-square (r.m.s.) variability in the QPO is $\geq 2.8\%$ and $\sim4\%$  for \suzaku\ and \x, respectively (Fig. S2).

\label{MONTECARLO}
Assuming  the signal is on top of a purely Poisson-noise time series, the limit on the r.m.s. variability found here results\cite{LewinvanParadijs1988SS} in a single-trial significance in the Gaussian limit of $\>3.8\sigma$ for \suzaku\ and $\>2.2\sigma$ for \x. However, in order to rigorously quantify the strength of the signal outside of the Gaussian regime and fully account for the presence of missing and unequally spaced data (Fig. 1), we conducted Monte Carlo simulations. The method we adopted is based on well established procedures in timing studies of compact objects\cite{LewinvanParadijs1988SS,vanderKlis1989,UttleyMcHardy2002MNRAS}. A series of 50,000 light-curves with the same average intensity, standard deviation, and number of bins as the original data were produced, and the noise power distribution was found at all Fourier frequencies and compared to the real observations (Fig. S4). In this manner, we found that the chance that the individual detections at $\sim5$~mHz  are due to random noise to be $P_{false{\rm |suzaku}}=~1.4\times10^{-4} $ and $P_{false{\rm |xmm}}  <5\times10^{-4} $ for \suzaku\ and \x\ respectively (Fig. S7).   After correcting  for the initial  ``blind-search"  of the \textit{Suzaku} data and accounting for the fact that the QPO was detected  in two independent observations, with different telescopes, the probability of two chance $3\sigma$ detections arising from random noise alone was found to be  $ 1.52\times10^{-5}$. The observed quasi-periodic signal at $\sim5$~mHz in \j\ is statistically highly significant ($4.33\sigma$ assuming Gaussianity in the probability).

QPOs are regularly seen in stellar mass black holes. Recently, QPOs have also been observed in a single AGN\cite{agnqpo2008}, and in a couple of potential intermediate-mass black holes ($\sim10^3~\msun$)\cite{ulxqpo22003,ulxqpongc50482007}. Despite the lack of a unique physical explanation, most models\cite{Nowak1997,CuiZhangChen1998,PsaltisBellonivanderKlis1999,StellaVietriMorsink1999,McKinneyTchekhovskoy2012} strongly link the origin of QPOs with orbits and/or resonances in the inner accretion disc close to the black hole. The detection of a QPO approximately 10 days after \j\ became active requires that an accretion disc was formed  shortly after the start of the initial TDF. The characteristic time\cite{rees1988,Strubbe2009}, $t_{\rm fall}$,  it takes for material from the disrupted star to fall back towards the black hole in \j\ from a pericenter distance of $R_p\sim13(M_{\rm BH}/10^6\msun)^{-5/6}~R_{\rm S}$\cite{Zauderer2011Natur} (where the Schwarzschild radius $R_{\rm S} = 2GM/c^2$) is $t_{\rm fall}\approx 1~{\rm day}$. This is consistent with the formation of an accretion disc due to stream-stream collision\cite{rees1988} after a small multiple of the $t_{\rm fall}$.  The Keplerian frequency at the radius of the innermost stable circular orbit (ISCO~=~$0.5-3R_{\rm S}$, depending on whether the black hole is spinning or not), just short of the black hole's event horizon, is generally the highest characteristic variability frequency. If the $5$~mHz QPO centroid frequency is set by orbits at the ISCO, it would imply a black hole mass between $\sim4.5\times10^5$ and $5\times 10^6\msun$ - for a non-rotating and maximally rotating black hole, respectively. This is in line with predictions based on simultaneous X-ray and radio observations\cite{millerkayhan2011} ($\sim3.2\times10^5\msun$) as well as the upper limit imposed by the $M-L$ relation ($<2\times10^7\msun$)\cite{Burrows2011Nature,Bloom2011Sci}.  Keplerian frequencies at the ISCO scale inversely with black hole mass; assuming a $\sim10^6\msun$ black hole,  the $\sim5$~mHz QPO would correspond to a frequency of  $\sim 500$~Hz around a $\sim10\msun$ black hole. This is  remarkably similar to the 450~Hz oscillation found for GRO~J1655-40\cite{Strohmayer2001}. Our results thus confirms a fundamental aspect of disc/disruption theory\cite{rees1988} and highlights the scale invariant nature\cite{fundamentalplane, fundamentalplane2, McHardy2006, gultekin2009} of the underlying physics governing the accretion flow onto supermassive ($\sim10^{5-9}\msun$) and stellar-mass ($\sim10\msun$)  black holes, several of which have displayed similar X-ray periodicities \cite{vanderKlisbook}.

Moreover, the Eddington limit for this black hole is  $<6\times10^{44}\ergps$, making the peak luminosity where the QPO is observed highly super-Eddington (Fig.~1). Thus, quasi-periodicities should be preserved in not only mildly  super-Eddington accretion flows, as is the case for the only other supermassive black hole  RE~J1034+396\cite{agnqpo2008,agnqpo22010} showing a QPO, and the stellar-mass black hole GRS~1915+105\cite{Remillard06}, but also in potentially highly beamed, anisotropic sources. Recent numerical simulations have began to examine the role of relativistic jets in the  production of QPOs\cite{McKinneyTchekhovskoy2012,Lei2012}, however,  even in this scenario, it is the connection between the accretion disk and the base of the jet that give rise to the quasi- periodic signal.

Although we have drawn from a rich literature that has ensured a high standard of robust techniques, there is still the concern that other uncertainties in the theoretical framework of accretion flows could be at play and result in unaccounted systematic errors in  the absolute noise model around black holes. Noise may not be purely white, purely red or a simple combination of the two, resulting in a smooth, featureless continuum. Indeed, \cite{Schnittman2006}, using general relativistic, magnetohydrodynamic simulations of the accretion flow onto a Schwarzschild black hole  showed possible high frequency quasi-periodic features which were identified with properties of the turbulent flow. If this turns out to be the general case, QPOs similar to the one detected here would still be highly useful in determining the physics of the accretion flow, but might not be trivially related to the fundamental properties  of the black hole, i.e.  mass and spin.

\begin{figure}[!t]
\label{Figure LCMAIN}
\begin{center}
\includegraphics[width=12.5cm]{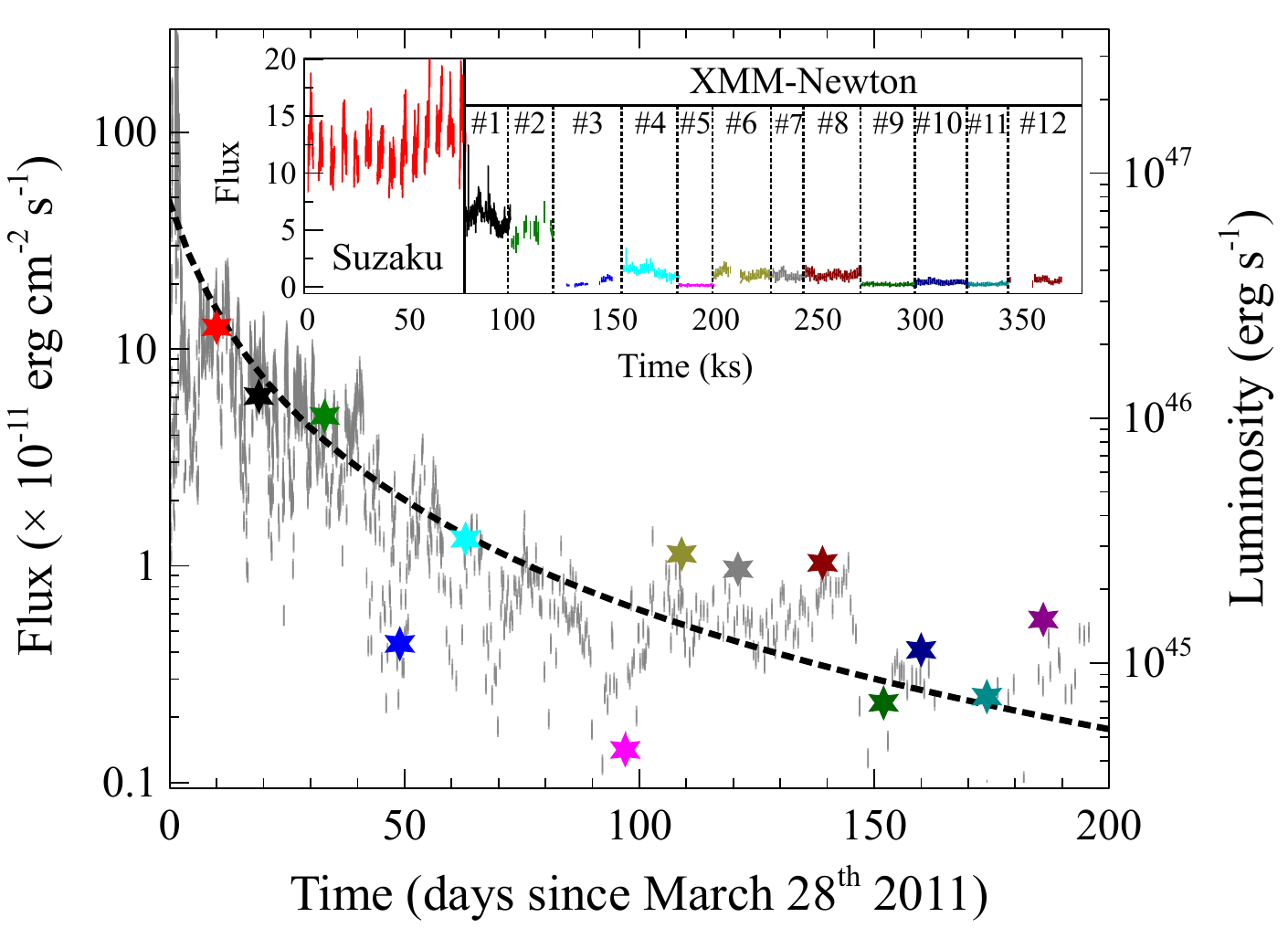}
\end{center}
\setlength{\baselineskip}{15.3pt}\textbf{ Fig. 1.} {\xmm\ and \suzaku\ light curve of Sw~J1644+57} together with the \swift-XRT 0.3-10\kev\ light curve for reference (grey)\cite{Evansswiftlc1,Evansswiftlc2}. For  twelve \xmm\ observations (\x-12)  we extracted 0.2-10\kev\ source and background light curves from the PN camera, using 40-arcsec circular regions. After accounting for the flaring background, a substantial fraction of the data in \textit{XMM}~\#2, 3 and 12 had to be excluded.  For \suzaku, we used a box region of 250-arcsec to extract  the source light-curve  from the two front-illuminated cameras and 150-arcsec for the background.  Every observation had a background level significantly less than 5\% of the total flux.  For each observation we created an energy spectrum to which we fitted a model consisting of an absorbed power-law and used this to obtain a conversion factor between the count-rate and fluxes in physical units. The average flux levels in each observation are shown in real time in the main axis (stars, where the one s.d. error bars are smaller than the symbol size)  and  the inset compares each observation with data points binned in 32-s intervals. The vertical lines separate the various observations. The right-hand axis gives the conversion to luminosity of the source assuming isotropic radiation and the dashed curve shows a $t^{-5/3}$ relation with $t$ being the time since March 28 2011. 
    \end{figure}

\begin{figure}
\begin{center}
\includegraphics[width=7.5cm]{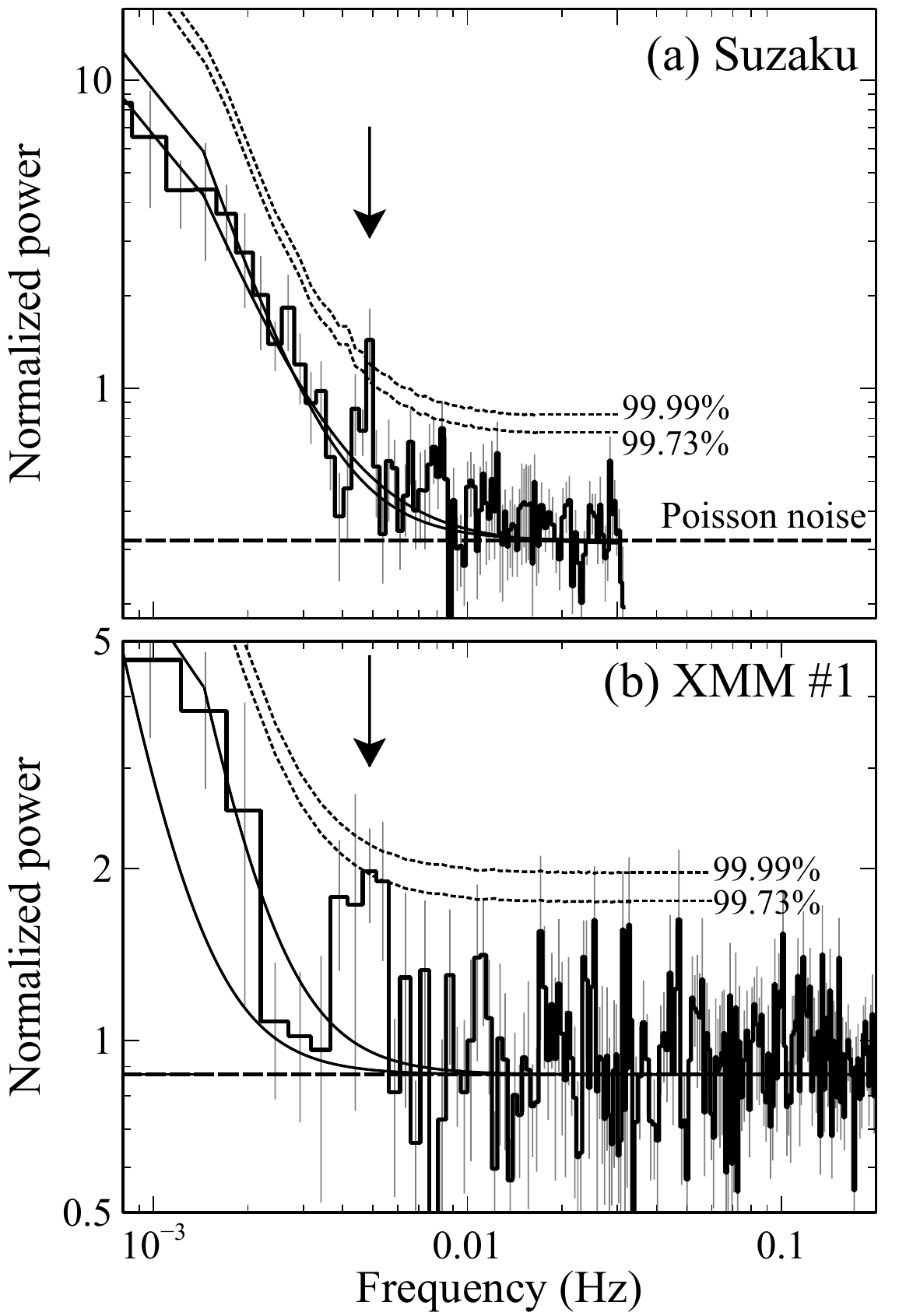}

\end{center}
\vspace*{0cm}
\setlength{\baselineskip}{15.3pt} \textbf{Fig. 2a} (Top): Power spectra for \suzaku\ and  \textbf{Fig. 2b} (Bottom): for \x, in the 2-10\kev\ energy range.   The power spectra are normalised such that their integral gives the squared r.m.s. fractional variability. The Poisson noise level  expected from the data errors is shown as the dashed horizontal line. We checked that the peak is robust to a variety of frequency and time resolutions.  The arrow in both panels mark the  presence of a QPO with a centroid frequency of  $\nu_{\rm suzaku}\sim4.8$~mHz. The solid curves enclose the range of the best  fit to the underlying continuum as  described in Table S1 (Model 2). The dotted curves show the 99.99\% and 99.73\% ($3\sigma$) threshold for significant detection. 
\end{figure}


\bibliography{scibib}

\bibliographystyle{Science}


\begin{scilastnote}

\item[Acknowledgements] R.R. thanks the Michigan Society of Fellows and NASA for support through the Einstein Fellowship Program, grant number PF1-120087. R.R. also wish to thank C.~Reynolds and R.~Mushotzky for comments on our early work. We all thank N. Schartel and the XMM-Newton staff for executing monitoring observations of Swift~J164449.3+573451. This work is based on observations made with XMM-Newton, a European Space Agency (ESA) science mission with instruments and contributions directly funded by ESA member states and the USA (NASA) and the Suzaku satellite, a collaborative mission between the space agencies of Japan (JAXA) and the USA (NASA). This work also made use of data supplied by the UK Swift Science Data Centre at the University of Leicester.

\item[Supplementary Materials] ~ \\www.sciencemag.org \\ Materials and Methods  \\Figs. S1 to  S9   \\References (37-56)

\end{scilastnote}

\clearpage

\label{Supplementary}
\begin{center}

{\title{\Large Supporting Online Material}} 

\vspace*{0.5cm}
\textbf{\title{\Large A 200-s Quasi-Periodicity Following the Tidal Disruption of a Star by a Dormant Black Hole}}
\end{center}

\begin{center}

\author{R.~C.~Reis,$^{1\ast}$  J.~M.~Miller,$^{1}$ M.~T.~Reynolds,$^{1}$  K.~G\"{u}ltekin,$^{1}$ D.~Maitra,$^{1}$ A.~L.~King,$^{1}$  \\ T. E. Strohmayer$^{2}$}
\end{center}

\begin{center}

{$^{1}$Department of Astronomy, University of Michigan, Ann Arbor, Michigan 48109, USA }

{$^{2}$Astrophysics Science Division, NASA Goddard Space Flight Center,}\\
{Greenbelt, MD 20771, USA }\\

\end{center}

\begin{center}

\vspace*{-.5cm}
{$^\ast$To whom correspondences should be addressed. E-mail:  rdosreis@umich.edu.}
\end{center}

\date{}


%
\section{Data Reduction} \label{dataReduction}
\subsection{\suzaku} \label{PSsuzaku}

{Suzaku} observed  Swift~J164449.3+573451 (hereafter \j) on 2011 April 6 for  a total ontime exposure of 52\ks. The three operating detectors constituting the X-ray Imaging Spectrometer (XIS\cite{SUZ_XIS}) on-board of \suzaku\ were operated in the normal, full-frame imaging mode with both front and back illuminated detectors in the 3x3 and 5x5 editing modes. Using the latest  \heasoftv\ software package we began by processing the unfiltered event files for each CCD following the \suzaku\ Data Reduction  Guide\footnote{http://heasarc.gsfc.nasa.gov/docs/suzaku/analysis/}. Detailed good time intervals (GTIs) were produced using the \ftool\ \xistime\ and setting the option  ``bstgti=no". New attitude files were then created using the script \aeattcor\footnote{http://space.mit.edu/cxc/software/suzaku/aeatt.html}  in  order to correct for shift in the mean position of the source caused by the wobbling of the optical axis\cite{aeattcor},  however, we note that this effect was minimal and the correction was only used for consistency.  The \ftool\ \xiscoord\ was used to create new event files which were then further corrected by re-running the \suzaku\ pipeline with the latest calibration, as well as the associated screening criteria files. The good time intervals mentioned above were also employed in all cases to exclude  any possible telemetry saturations.  \xselect\ was used to extract spectral products from these event files with light curves being created with 16 seconds time resolutions -- the minimum possible in this observing mode. The final good time exposure was 38.1\ks.

In order to estimate the level of pile up suffered by the data we used the  script\cite{pile_estimate} {{\rm{\small PILE\_EST}}}\footnote{http://space.mit.edu/cxc/software/suzaku/pest.htm}  to create a pileup map out of a \suzaku\ event data file. It was found that the data suffers very little from pile up (maximum pileup fraction of $5\%$) and therefore we chose to employ a square box  region with a  width of  240~pixels ($\sim250$'') without excluding any data. Background spectra were extracted from a box region having a width of 150"  elsewhere on the same chip.  Individual ancillary response files (arfs) and redistribution matrix files (rmfs) were produced with the script {{\rm{\small XISRESP}}}{\footnote {http://suzaku.gsfc.nasa.gov/docs/suzaku/analysis/xisresp}} -- which calls the tools {{\rm{\small XISRMFGEN}}} and {{\rm{\small XISARFGEN}} -- with the ``medium''  input. The resulting spectra had a 0.5-10\kev\ count rate of $\sim4.3$ and $\sim4.1$c/s for the front illuminated and back illuminated detector respectively, with the background constituting $\sim0.4$\% and $\sim0.6$\% of the total flux.

\subsection{\xmm} \label{dataxmm}

{\it XMM-Newton} performed twelve, approximately bi-weekly pointing observations of \j\ starting on 2011 April 16 and finishing 2011 October 3, with each observation lasting between  $\sim16$\ks\ to $\sim26$\ks. In all cases, the \epicpn\ camera\cite{XMM_PN} on-board \xmm\ was operated in ``imaging'' mode with a ``medium'' optical blocking filter. The \epicmos1\ and \epicmos2\ cameras\cite{XMM_MOS} were operated in the ``imaging'' data mode with the  \epicmos1\ and \epicmos2\ being in the ``PrimePartialW2" and ``FastUncompressed" submodes respectively. In this work we only use the \epicpn\ data, from which we always start with the unscreened level 1 data files. We generate concatenated and calibrated event lists using the latest \xmm\ {\it Science Analysis System \thinspace v 11.0.0 (SAS)}. \epicpn\ events were extracted from a circular region with approximately 40" but varying slightly depending on how far the observation is from the chip gap. Bad pixels and events too close to chip edges were ignored by requiring ``FLAG = 0'' and ``PATTERN $\leq$ 4''. Flaring background meant that a significant fraction of of the data in \textit{XMM}~\#2, 3, and 12 had to be discarded. The energy channels in all observations were initially binned by a factor of five to create an energy spectrum and both source background light curves were produced in the 0.2-10\kev\ energy range. We used the SAS task \epatplot\ to assess the level of pile-up in the \epicpn\ spectrum and found it to be insignificant. The highest \epicpn\ count rate is approximately 10~\ctsps\ for \x,  which is below the nominal pile-up limit for the instrument in ``LargeWindow" submode. Response files were created in the standard way using the tools \rmfgen\ and \arfgen. The total good exposure time for \x\ was 16.96\ks.  Finally we rebinned the energy spectrum with the tool \pharbn {\footnote{http://virgo.bitp.kiev.ua/docs/xmm$_{-}$sas/Pawel/reduction/pharbn}}, to have 3 energy channels per resolution element, and at least 20 counts per energy channel.

\section{Energy Spectra} \label{dataAnalyses}

\subsection{\suzaku} \label{ESsuzaku}

\begin{figure}[!t]
\label{fig1}
\begin{center}
\hspace*{-1.cm}
\includegraphics[angle=0, width=7.8cm]{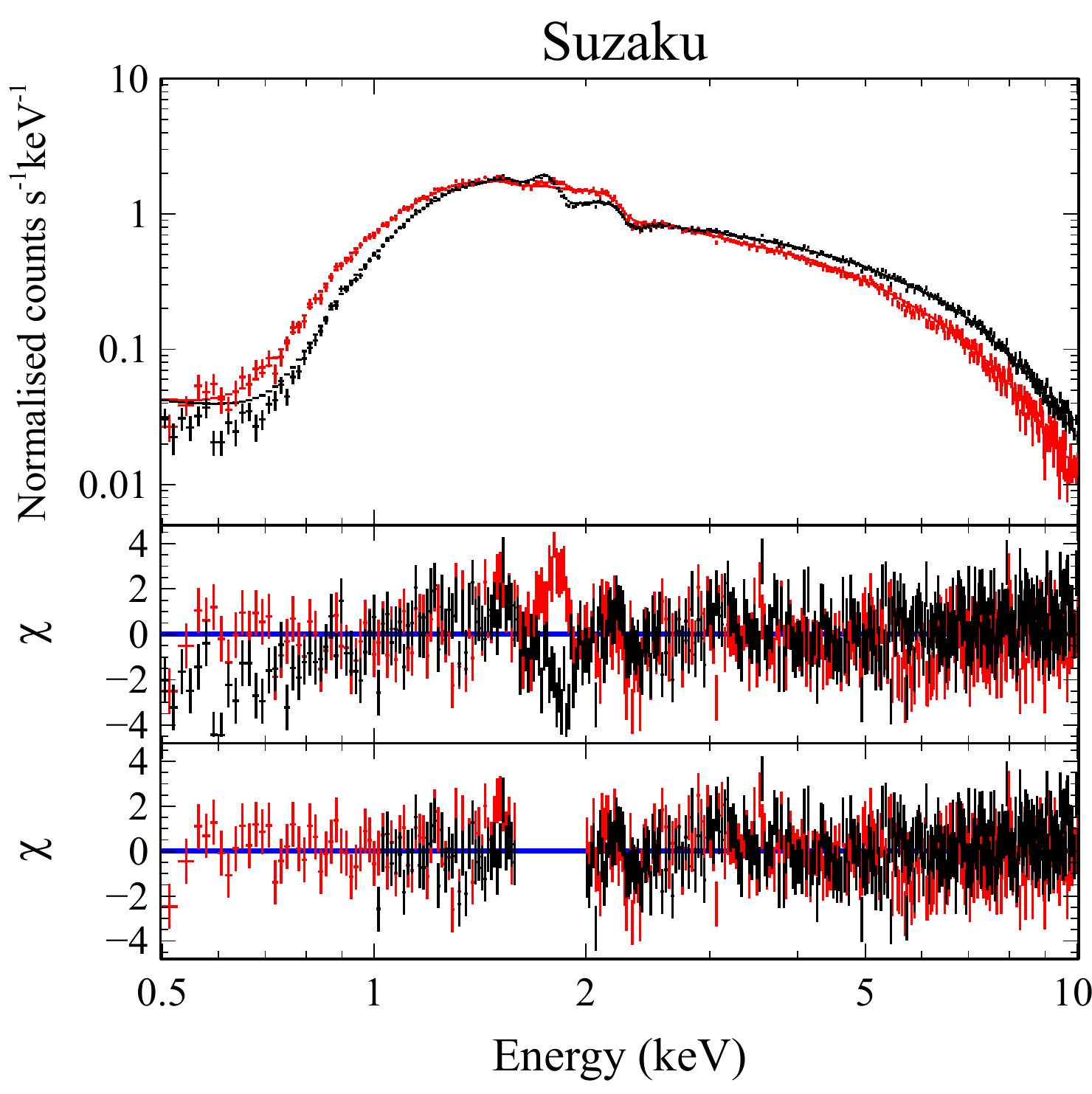}
\hspace*{-0.cm}
\includegraphics[angle=0, width=7.8cm]{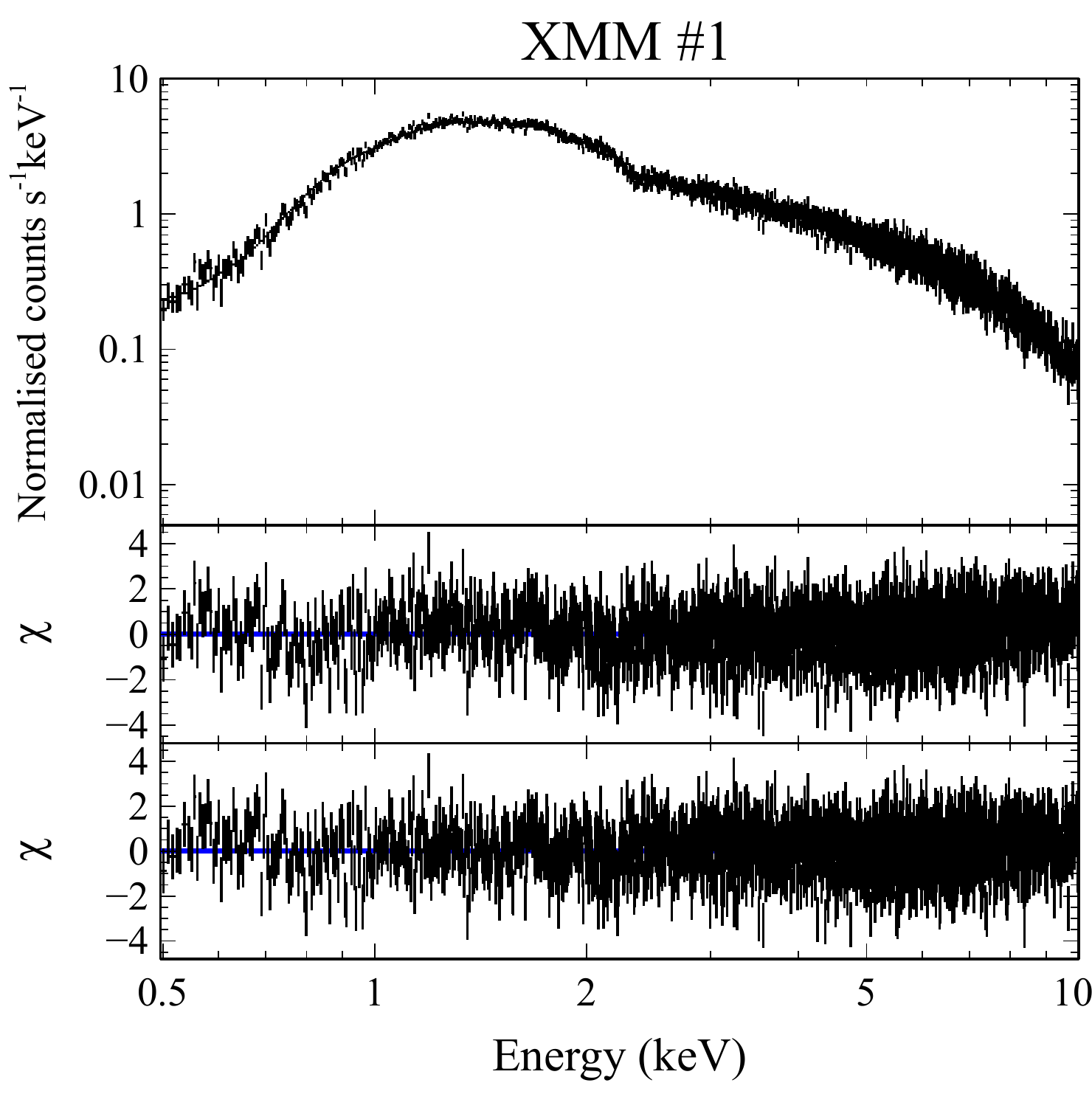}
\end{center}
\vspace*{-0.5cm}
\setlength{\baselineskip}{15.3pt} {\bf Fig. S1a} (left): Front  (black) and back (red) illuminated \suzaku\ spectra (top) fit with an absorbed powerlaw.  The  middle panel shows the residual to this fit over the full energy range and the bottom panel after excluding the silicon region, the front illuminated data below 1\kev\ and adding a phenomenological disk black body.  {\bf Fig. S1b} ( right) Same for the \textit{XMM}~\#1 \epicpn\ spectrum in the full energy range.  Note the severe drop in total counts below $\approx 1$\kev\  as well as the clear presence of un-calibrated silicon features just below 2\kev\ in the \suzaku\ data.
\end{figure}

We analysed briefly the energy spectra within the X-ray fitting package \xspec\cite{XSPEC}. A model consisting of an absorbed power law (\zphabs$\times$\zpo\ in \xspec)  resulted in a  fit with $\chisq/{\rm D.o.F} = 1181.2/779$ (Reduced-$\chisq = 1.516$) which is not formally acceptable. However, as can be seen from Fig.~S1a (top and middle panels), quite a number of the systematic residuals are in fact coming from the region between  $\sim 1.6 -– 2.0 $\kev,  likely due to the known systematics uncertainties in the effective area near the silicon K and gold-M edge.  Excluding this region  improves the fit to $\chisq/{\rm D.o.F} = 982.6/723$ (Reduced-$\chisq = 1.36$) with now the only noticeable systematic residues coming from the front illuminated data below $\sim 1$\kev\ as a result of its smaller effective area as compared to the back illuminated camera. Excluding the FI data below 1\kev\  results in $\chisq/{\rm D.o.F} = 833.9/688$ (Reduced-$\chisq = 1.21$). 

\swift\ observations of \j\ during the first two days after its detection showed that the system exhibits times where the spectrum is well fit by an absorbed powerlaw at comparably low-flux states and preferably a black body at higher-flux states\cite{Bloom2011GCN}. In fact, the authors suggest that during those observations, the spectra are better fit by a superposition of a thermal disk with a temperature of $\sim 1$\kev, and a non-thermal component. A similar study using the combined \swift-XRT data for the first 10 days after the trigger\cite{Burrows2011Nature}  finds that the spectra at various intensity levels are comparably well fit by either an absorbed log-parabola model, absorbed broken-powerlaw model or an absorbed powerlaw model plus an extra soft thermal component. However, a simple absorbed powerlaw model was inferior to the other combination during the first 10 days. The authors also find that during that time, the spectra hardens from $\Gamma\sim2.3$ to $\Gamma\sim 1.8$  as the intensity increases. A further study\cite{Levan2011Sci}  also explore the \swift-XRT  data during the first $\sim15$ days and find a time-averaged photon index of $\Gamma=1.80\pm0.25$ when modelling with an simple absorbed powerlaw, but also states that  a powerlaw with an index of $\Gamma = 1.6$ having an exponential cutoff at $\sim$1\kev\ can also model the data equally well. This latter phenomenological combination has the same effect as an extra soft component affecting the spectra below $\sim1$\kev, similar to the phenomenological addition of a thermal disk component used by \cite{Bloom2011GCN,Burrows2011Nature}.

 We explore whether the addition of a \diskbb\ component significantly improve the quality of the fit and find that it provides only a mild improvement of  $\Delta\chisq = 27.7$ for 2 degrees of freedom. This fit (shown in the bottom panel of Fig.~S1a) has an absorbing column \nh\ of\footnote{All errors shown in this section are 90 per cent confidence errors obtained by allowing all parameters to vary. } $(1.81\pm0.06)\times 10^{22}\pcmsq$, a powerlaw index of $1.83\pm0.02$ and a disk temperature of $0.35^{+0.11}_{-0.06}$\kev. The unabsorbed 0.5-10\kev\ powerlaw and \diskbb\ fluxes are $1.93^{+0.02}_{-0.04} \times10^{-10}\ergpcmsqps$ and $8^{+4}_{-3} \times10^{-12}\ergpcmsqps$, respectively.

It should be stressed that it is often the case with active galactic nuclei that energies below a few \kev\ show the presence of a broad and smooth ``soft excess"  (see e.g. \cite{Crummy06}) which,  despite the observed AGN hosting black holes spanning many orders of magnitude in mass, remains rather constant in temperature when fitted with a phenomenological disk black body (e.g. \cite{gierlinskidone04}) suggesting that the excess is in fact atomic in nature and not the true disk emission which often peaks in the UV. As such, the temperature quoted above is only a fiducial disk temperature and should not be though of as the actual accretion disk temperature.

\subsection{\xmm} \label{ESXMM}

We find that a model consisting of an absorbed powerlaw (\zphabs$\times$\zpo\ in \xspec) provides a formally acceptable  fit to all twelve \xmm\ spectra, with reduced-$\chisq$ ranging from 0.9 to 1.14.  Simultaneous fit to all 12 spectra, allowing for variation in all model parameters between observations,  resulted in $\chisq/{\rm D.o.F} = 8190.3/8219$ (Reduced-$\chisq = 0.997$).  The data shows a distinct hardening of the powerlaw index with time with the power law index going from $1.72\pm0.01$ in the first observation (day $\sim19$) to $1.52\pm0.02$ for \textit{XMM}~\#2 (day $\sim33$ and settling at approximately 1.3 at later observations (for \textit{XMM}~\#6-12, $\Gamma<1.4$). Figure 1 in the main journal shows the evolution in the flux level of the twelve \xmm\ observations, together with the preceding \suzaku\ pointing as well as the \swift-XRT light curve for reference.

The total column density varies in the range $\sim (1.32-1.62)\times 10^{22}\pcmsq$ and assuming the spectra are indeed  simple powerlaws, we  find that this variation is statistically significant, in contrast with \cite{Levan2011Sci}. A model having a static value for \nh\ of   $(1.41\pm 0.01) \times 10^{22}\pcmsq$ yields $\chisq/{\rm D.o.F} = 8441.9/8230$  (Reduced-$\chisq = 1.03$).  However, if we add a phenomenological soft-excess in the form of a \diskbb, an equally valid fit ($\chisq/{\rm D.o.F} = 8175.8/8206$;  (Reduced-$\chisq = 0.996$) can be obtained with a static \nh\ and variable disk flux and temperature.  In the first five \xmm\ observations where the temperature of this fiducial soft-excess is well constrained\footnote{Defined here as having errors less than 50\% its value.}, we find its value to range between  0.24--0.36\kev\ (compare with the \suzaku\ value of $0.35^{+0.11}_{-0.06}$\kev).  Fig.~S1b shows the \epicpn\ spectrum of \x\ fit  with a single powerlaw (top and middle panels) as well as   after the inclusion of a \diskbb\ (bottom panel).  The smoother decline in the soft flux of \x\ as compared to \suzaku\ is likely due to the lower \nh\ of $(1.36\pm 0.04) \times 10^{22}\pcmsq$ compared to $(1.81\pm0.06)\times 10^{22}\pcmsq$ for \suzaku.

\section{ Power Spectrum }

\subsection{\suzaku} \label{PSsuzaku}

\begin{figure}
\label{fig2}
\begin{center}
\hspace*{-0.cm}
\includegraphics[width=8.cm]{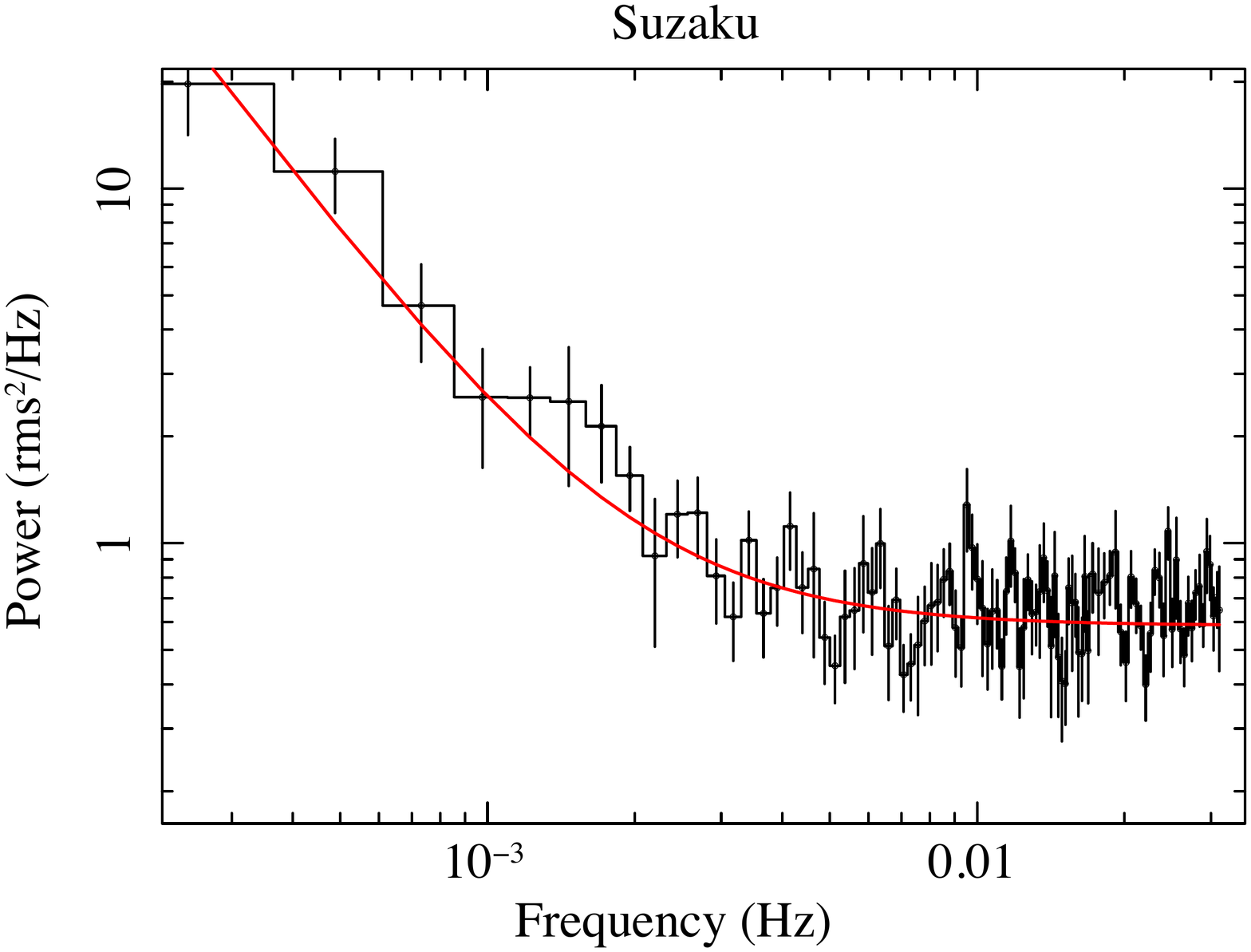}
\hspace*{-1.cm}
\includegraphics[width=8.cm]{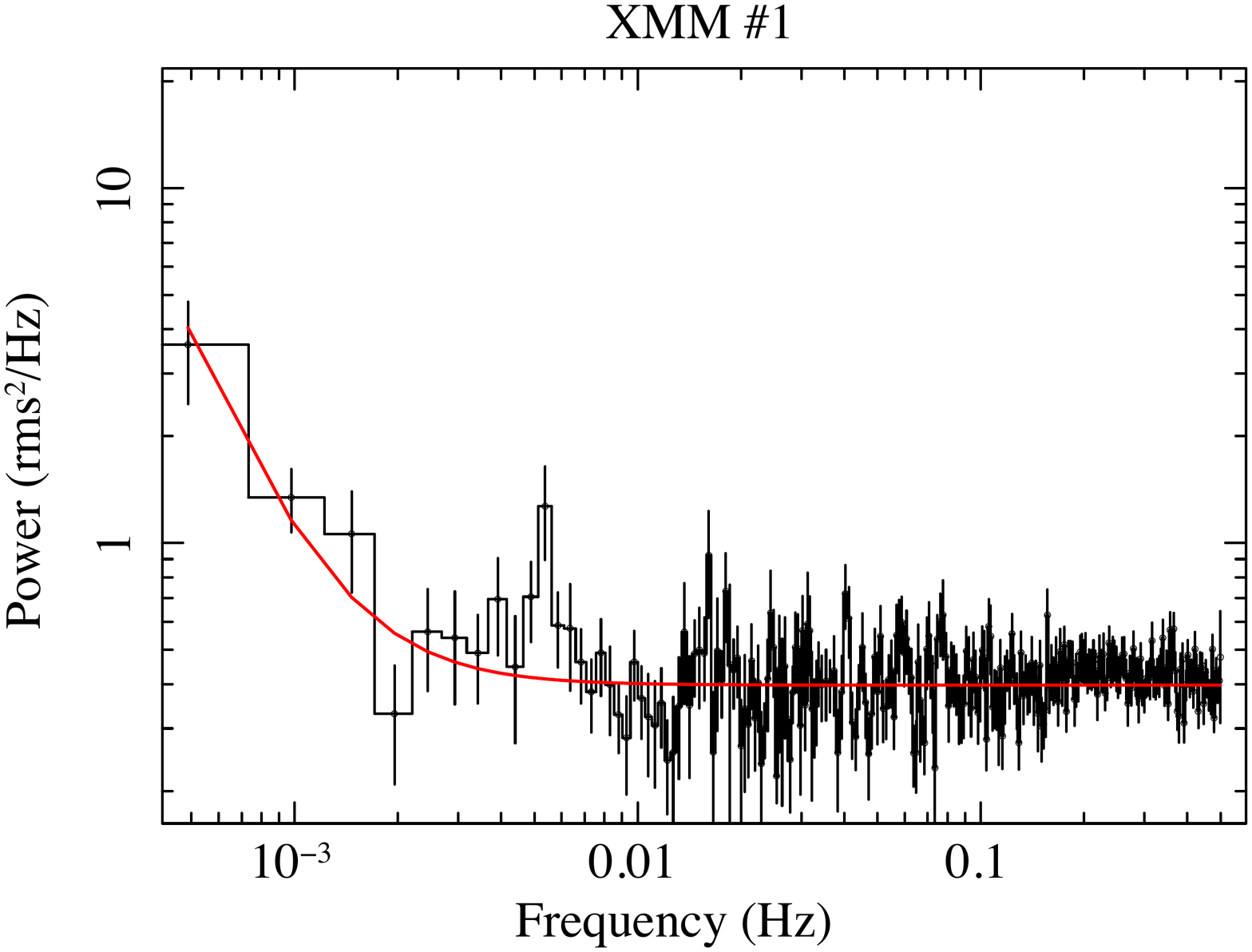}
\end{center}
\vspace*{-1.cm}
\setlength{\baselineskip}{15.3pt} {\bf Fig. S2a} (left): Average power spectrum for \textit{Suzaku}  and  for \textit{XMM}~\#1  {\bf Fig. S2b} (right) obtained for energies below 2\kev. Note the differing frequency  ranges. 

\end{figure}

\suzaku's low orbit means that most targets are occulted by the earth for about a third of its $\sim5.8$\ks\ orbit. For this reason, the  38.1\ks\ observation in fact consists of fourteen periods ranging from $\sim1.98$\ks\ to $\sim3.0$\ks. To create a combined power spectrum, we broke the full observation into these fourteen continuous time series each having identical frequency bins and averaged their power spectra. We used 256~s intervals to produce a total of 99 frequency bins ranging from $2.44\times10^{-4}$~Hz to  $3.125\times10^{-2}$~Hz with frequency resolution of $1.22\times10^{-4}$~Hz up to $1.68\times10^{-2}$~Hz and twice that at higher frequencies (geometrically binned) in order to produce the individual power spectra.  A visual inspection of the averaged 0.2-2.0\kev\ power spectrum  displayed no distinct, narrow features with a model consisting of Poisson noise and a  power law component yielding a best-fit (least-squares method) of $\chisq/\nu = 115.2/96$ and $\Gamma=-1.82\pm0.19$ ($\Delta\chisq=2.71$ for 1 D.o.F). This is shown in Fig. S2 (left).  However, the spectrum above 2\kev\ (Figure 2 of the main journal)  shows a significant feature at around 5mHz. Visual inspections of power spectra having twice and four times the frequency resolution also indicates the presence of this feature. A fit with the same model as to that for the spectrum below 2\kev, i.e. a constant together with a power law,  results in a  fit with $\chisq/\nu = 156.1/96$ (Model 1;  Table~S1) and an index of $\Gamma=-1.79\pm0.12$. Adding a break to this power law so that below a frequency $F_{break}$ the continuum is described as $\Gamma_1$ and above as  $\Gamma_2$  resulted in a significant  improvement to the continuum  ( ($\chisq/\nu = 136.0/93$; Model 2;  Table~S1). From Fig. S6 we can see the visual improvement of a broken power law over the simpler continuum.

An equally satisfactory description of the continuum ($\chisq/\nu = 140.6.1/94$) can be achieved by replacing the broken power law with the sum of two broad Lorentzian components\cite{Nowak2000qpo, Homan2003, Homan2005}, one having a centroid frequency frozen at zero and the other at a frequency similar to the break frequency above. In all cases, however, the presence of a feature above the continuum around 5mHz remains 	and we  finally model this QPO with a Lorentzian having a centroid frequency of $4.81^{+0.08}_{-0.03}$~mHz and a quality factor $Q_{Suzaku}=\delta\nu/\nu >12$.

\begin{table}
\label{table1}
{\textbf{Table S1:} Fits to the power spectra of  \textit{Suzaku} and \textit{XMM}~\#1.}    

\begin{footnotesize}
\begin{tabular}{ccccccc}
\hline 
• & \multicolumn{2}{c}{Model 1} & \multicolumn{2}{c}{Model 2} & \multicolumn{2}{c}{Model 3} \\ 
\hline 
• & \suzaku\ & \textit{XMM} \#1 & \suzaku\ & \x\ & \suzaku\ & \x \\ 
\hline 
Poisson level & $0.30\pm0.02$ & $0.872\pm0.014$ &$0.32\pm0.02$  & $0.872\pm0.014$ & $0.32^{+0.02}_{-0.01}$ & $0.871\pm0.014$ \\ 
 
$\Gamma_1$ & $-1.79\pm0.12$ & $-2.1^{+0.4}_{-0.6}$ & $-1.3^{+0.4}_{-0.3}$ & -1(f) & $-1.3^{+0.4}_{-0.3}$ & -1(f) \\ 
 
$\Gamma_2$ & --- & --- & $-2.90^{+0.50}_{-0.04}$ & -3(f) &  $-3.18^{+0.55}_{-0.03}$& -3(f) \\ 

$F_{break}$ (mHz)  & --- & --- &$1.4^{+0.4}_{-1.0}$  & $<1.7$ & $1.4^{+0.4}_{-1.0}$ & $<1.7$  \\ 
 
$N_{1}$ ($\times10^{-5}$) & $1.8^{+2.1}_{-1.0}$ & $0.095\pm0.037$ & $75\pm15$  & $330\pm140$ & $75\pm15$  & $330\pm140$ \\ 
 
$N_{2}$ ($\times10^{-9}$) & --- & --- & $31^{+500}_{-28}$  & $7\pm4$ &$6^{+150}_{-2}$ & $6\pm4$ \\ 
 
 $\nu$ (mHz) & --- & --- & --- & --- & $4.81^{+0.08}_{-0.03}$ & $4.7^{+0.5}_{-0.2}$ \\ 
 
$\delta\nu$(mHz) & --- & --- & --- & --- & $<0.4$ & $0.298^{+0.001}_{-0.029}$ \\ 
 
$N_{\nu}$ & --- & ---& --- & --- & $3.1\pm1.5$ & $3.6\pm2.0$  \\ 
 
$\chi^2/\nu$ & 156.1/96 & 392.5/250  & 136.0/93 & 384.9/249 & 125.6/90 & 375.9/246 \\ 
\hline 
\end{tabular}
\vspace*{0.2cm}

Note --  Model~1 consists of a power law together with Poisson noise. Model~2 introduces a break to the power law and Model~3 adds a  Lorentzian component to account for the presence of the QPO. In the case of \x, the two power law indices could not be constrained so we froze them at the values shown.
\end{footnotesize}

\end{table}

\subsection{\xmm} \label{PSxmm}

We produced power spectra for all \xmm\ observations, and the resultant power spectra were averaged together. The presence of a feature at a similar frequency to that shown for \suzaku\ above was immediately clear; however, by looking at the individual power spectra from each observation we found that the signal was driven by the first pointing. This is not surprising given the clear change in the spectral  properties of the source between \x\ and the latter observations mentioned above, and the closer resemblance \x\ has with the spectral shape of (and proximity in time with) the \suzaku\ observation. Indeed, QPOs are known to be transient features with a good example being the QPO found in the active galaxy RE~J1034+396 where the $5.6\sigma$ discovery\cite{agnqpo2008} was not seen in three follow up observations nor in a preceding  observation\cite{MiddletonUttley2011}. As such, we concentrate below on the description of the power spectrum made for \x.  Figure S2b shows a  power spectrum for energies below 2\kev\ fit with a power law with index $\Gamma=-2.26^{+0.51}_{-0.02}$ ($\chisq/\nu = 341.6/250$).  In contrast to \suzaku, the corresponding power spectrum shows visual sign of a feature at $\sim5$mHz even at this soft energy range.  

The lack of any significant feature below 2\kev\ in the \suzaku\ data can be understood in the context of the previous section in that not only does the total counts for \suzaku\ drops dramatically at these energies, it also suffers from clear calibration problems and confusion between possible variation in the column density or the presence of a further soft-excess component. As is clear from Fig.~S1, the spectrum of \x\ does not contain a comparatively strong drop in counts, partly due to its higher effective area as well as possibly lower column density or stronger soft-excess component at low energies, nor does the spectrum suffer from distinct calibration. These various factors propagates into a detection of a feature in the power spectrum below 2\kev\ in \x. 

As detailed in the previous section, we cannot break the degeneracy found at the low energy range between a model having a variable column density with a powerlaw continuum or a static column with a further variable soft-excess. With this in mind,  as well as to be consistent with \suzaku\ where a further difficulty arises due to low effective area and calibration uncertainties, we proceed through the remainder of the manuscript by examining the data above 2\kev. Figure~2 in the main journal shows the power spectrum at harder energies similar to that of \suzaku. We used  2048~s intervals to produce a total of eleven power spectra which were averaged to make a single power spectrum with 253 frequency bins ranging from $4.88\times10^{-4}$~Hz to  $4.998\times10^{-1}$~Hz with frequency resolution of $2.44\times10^{-4}$~Hz up to $3.37\times10^{-2}$~Hz and  increasing at higher frequencies (geometrically binned) in order to produce the individual power spectrum. A visual inspection of the averaged power spectrum shown in Figure~2 of the main journal  shows a significant feature at around 5mHz. Fitting a single power law to the continuum yields $\Gamma=-2.21^{+0.4}_{-0.6}$ with $\chisq/\nu = 392.5/250$. Replacing the single power law with a model similar to that of \suzaku, having $\Gamma_1$ and $\Gamma_2$ frozen at -1 and -3 respectively results in an improvement to the continuum with $\Delta\chisq/\Delta\nu = 7.6/1$ (Table~S1). A similar best-fit is obtained when replacing the broken power law with a combination of two broad Lorentzian  ($\chisq/\nu = 385.4/248$), again in line with the results presented for \suzaku\ above. Adding a Lorentzian to model the QPO  at $\sim5$mHz, we find that the QPO in the \x\ data has a centroid frequency of $4.7^{+0.5}_{-0.2}$~mHz and a quality factor $Q_{XMM} \approx 15$. Within the precision of the measurements, the frequency of the QPO appears to be stable between both independent observations.

\subsection{\swift-XRT Timing Analysis} \label{SWIFT}

Early results\cite{Burrows2011Nature} claimed to find no evidence for any statistically significant QPO in the \swift-XRT curve during the first twenty days of the outburst.  We have performed an analogous Fourier analyses on the data belonging to the first 20 days of the observations and summarise the results in Fig.~S3. Due to the random observing cadence employed by the XRT, the power spectrum is highly aliased and dominated by noise making the various apparent  features not statistically significant, in agreement with early results. Interestingly, although not statistically significant, if we zoom into a narrower frequency range centred at $\sim5$mHz (Fig.~S3b) we do see a potential feature which, when modelled with a Lorentzian on top of Poisson and red noise, results in a centroid frequency of $\approx6.2$mHz and a quality factor $Q\approx 20$.

\begin{figure}[!t]
\label{fig3}
\begin{center}
\hspace*{-0.5cm}
\includegraphics[width=8.5cm]{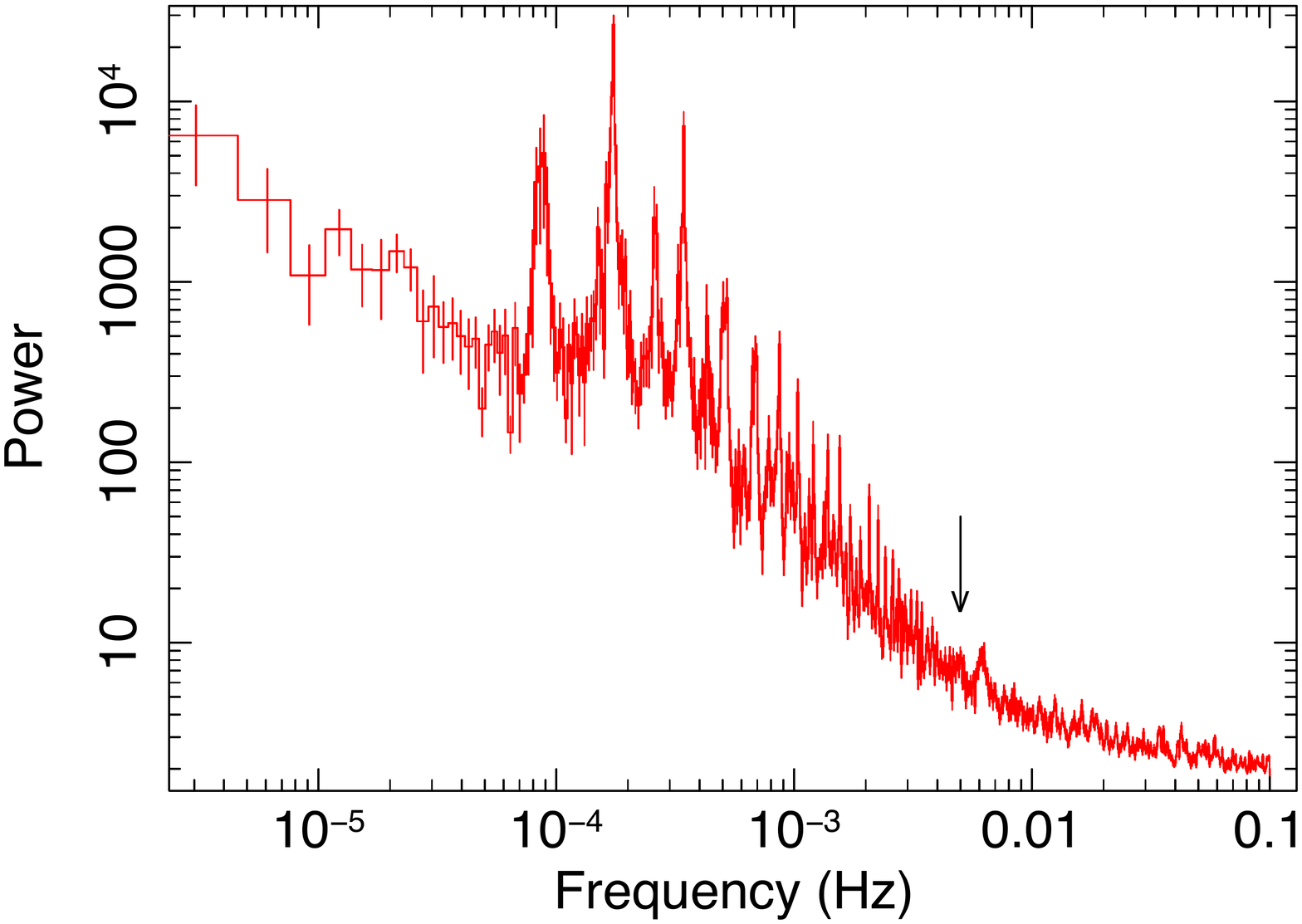}
\hspace*{-1.cm}
\includegraphics[width=8.5cm]{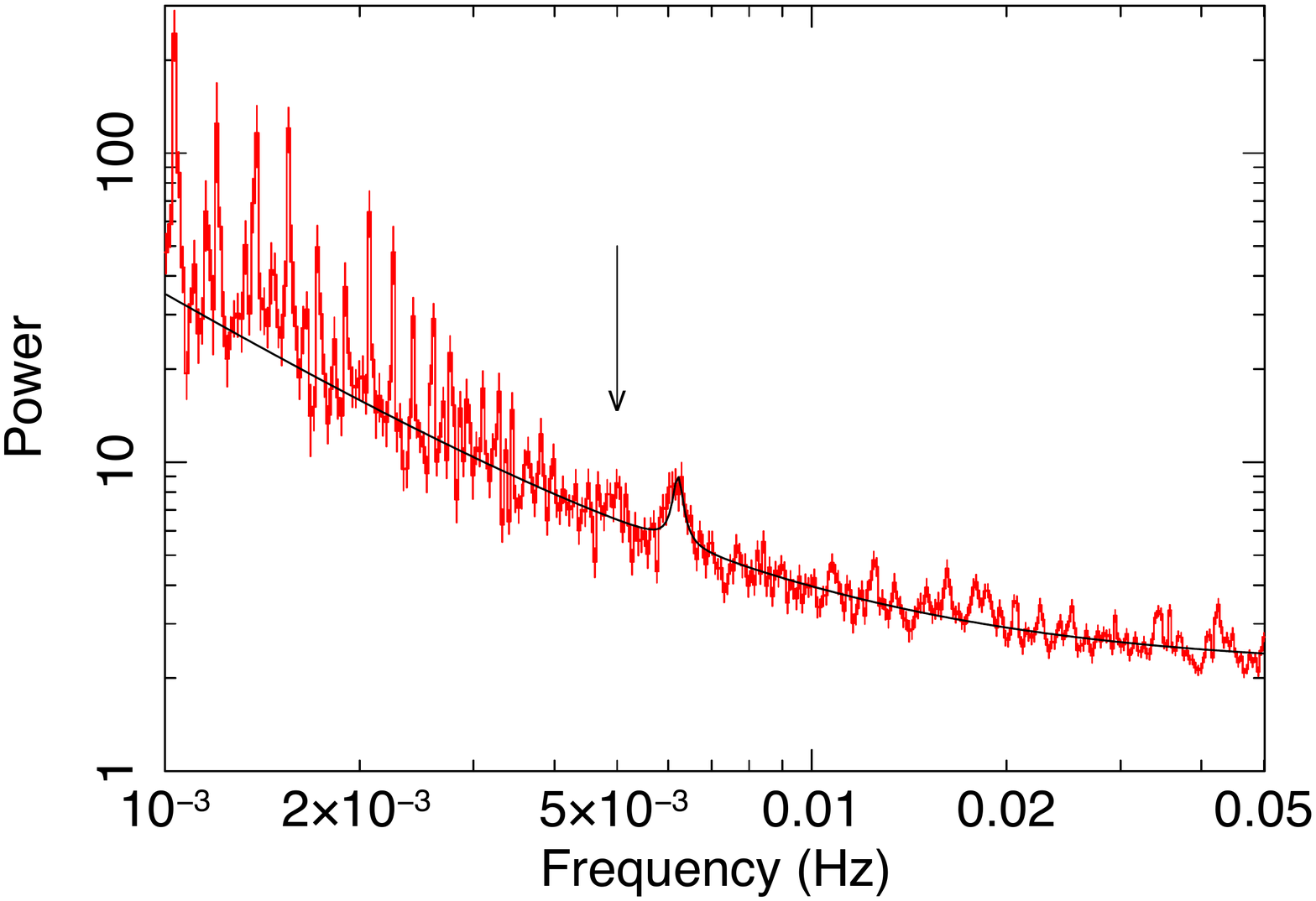}
\end{center}
\vspace*{-1.cm}
\setlength{\baselineskip}{15.3pt} {\bf Fig. S3a} (left): Swift-XRT power spectrum made from the combined data from the first $\sim20$ days after the outburst. The presence of various features at low frequencies are immediately obvious and are likely a result of  aliasing due to the random observing cadence employed by the XRT.  {\bf Fig. S3b} (right) Close up at higher frequencies showing the possible presence of a feature clearly above the local noise level, which, when fit with a Lorentzian, results in a centroid frequency of $\approx6.2$mHz and a quality factor $Q\approx 20$. We emphasised that due to the noise characteristics of this power spectrum, this feature is not statistically significant. The frequency of the QPO reported in this manuscript is marked by the vertical arrow.

\end{figure}

\section{Monte Carlo Sampling}\label{BOOT}

In order to confirm the accuracy of the estimated values presented in Table~S1, we proceeded by generating a sample of 10,000 light curves sampled by the empirical distribution of the observed light curves themselves. Let $c_i$ be the count rate and $c_{{\rm eer};i}$  the associated error of the $i^{th}$ time bin in the observed light curve. Based on this input light curve, 10,000 bootstrapped\cite{bootstrap} light curves were simulated using the following prescription:  the $i^{th}$ time-bin of a given simulated light curve have a count-rate of $cc_i$ and count-rate error $c_{{\rm err}; i}$ with  $cc_i$ being drawn from a normal distribution $N(c_i, c_{{\rm err};i}  )$.  The GNU Scientific Library (GSL) function \textit{gsl{\_}ran{\_}gaussian} was used to draw a Gaussian random variate with mean zero and standard deviation $c_{{\rm err};i}$. The value returned by \textit{gsl{\_}ran{\_}gaussian} was added to $c_i$ to obtain $cc_i$. Since observed fluxes cannot be negative, we set $cc_i$ to be equal to zero when its value returned by the above prescription was $<=0$. A similar approach was used to generate GRB light curves by Stamatikos et al\cite{Stamatikos2009} (see also \cite{Norris2000ApJ}).

For each light curve, we generated  power spectra in an identical manner to that described in the previous sections and modelled the data with the template described in Model~3,  with all parameters  except for the centroid frequency  of the Lorentzian describing the QPO  free to vary. Figure~2 shows the fractional root-mean-square (r.m.s) variability in the QPO  as found by the integral under the Lorentzian in the ``r.m.s. normalised" power spectra, ie. it is the square root of the product $\pi N_{\nu} \delta\nu/2$. The QPO in the \suzaku\ data is clearly very narrow (See Figure~2 in the main journal), possibly lying within just one frequency bin.  The upper limit shown in  Table~S1 give rise to the low peak in Fig. S4 at $\approx2.8$\%. We checked this limit on the fractional variability by further  running a series of simulations with the width of the Lorentzian  frozen at $4\times10^{-4}, 1\times10^{-5} ~{\rm and}~ 5\times10^{-6}~$Hz, and in all cases we obtained a lower limit consistent with the result shown in this figure.  For \x, we find that the r.m.s. variability in the Lorentzian is better constrained at $\approx 4\%$ compared to that of \textit{Suzaku}, in line with the constraints on the width and normalization quoted in Table~S1. 
 
\begin{figure}
\label{fig4}
\begin{center}
\hspace*{-0.6cm}
\includegraphics[width=8.cm]{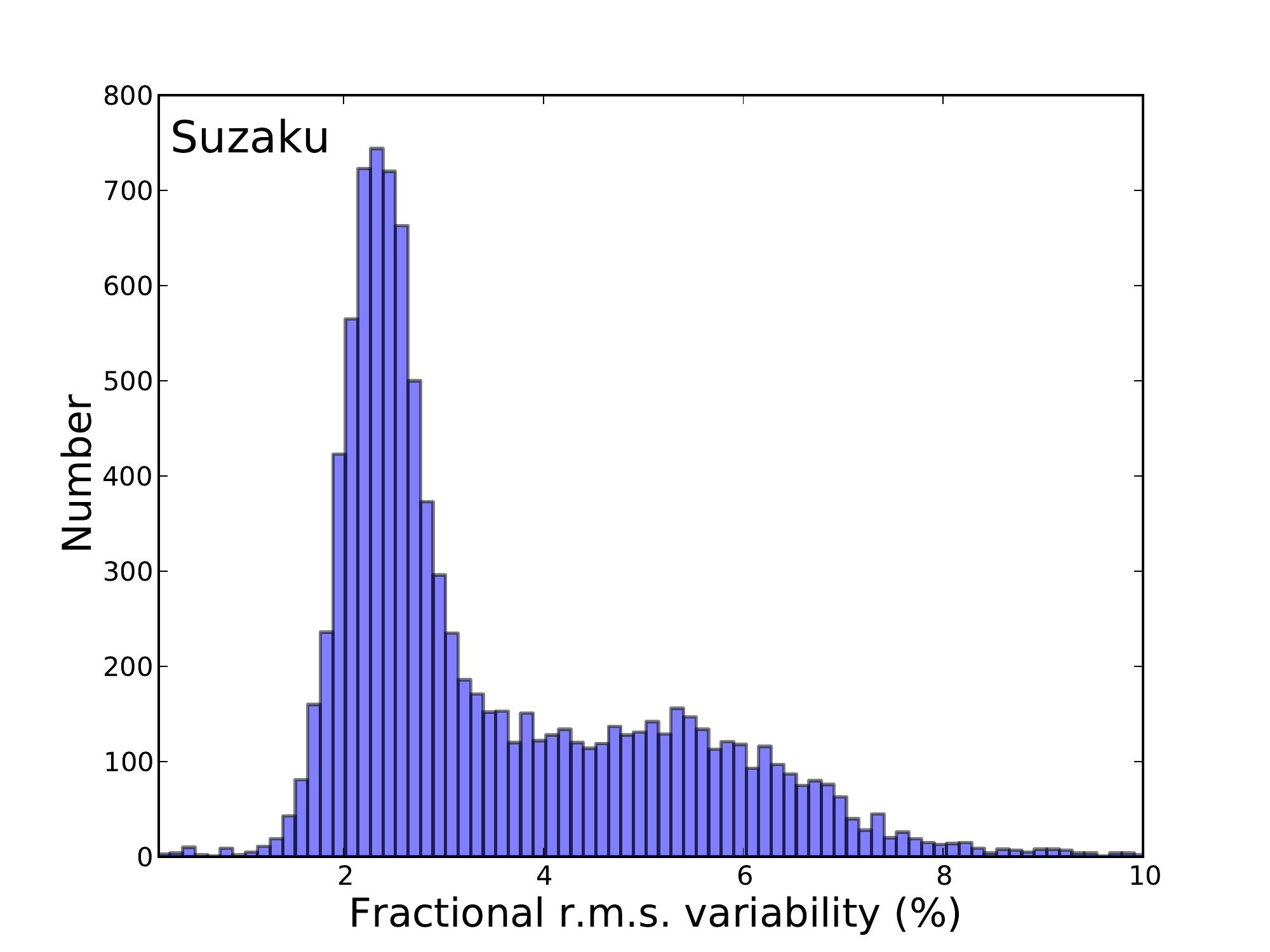}
\hspace*{-0.8cm}
\includegraphics[width=8.cm]{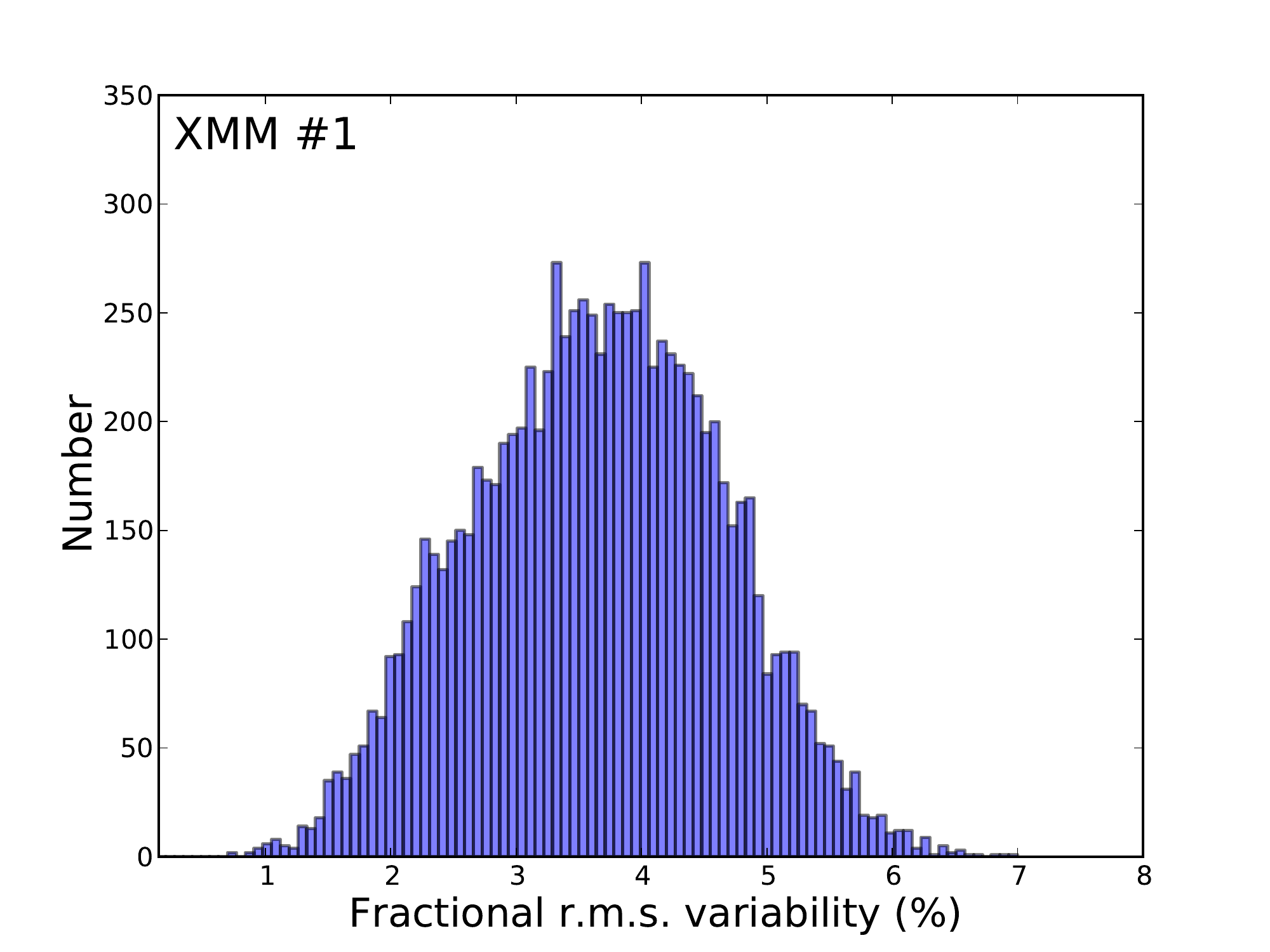}
\end{center}
\vspace*{-.5cm}
\setlength{\baselineskip}{15.3pt} {\bf Fig. S4a} (left): Fractional root-mean-square variability in the Lorentzian component modelling the QPO in 10,000 bootstrap samples for \suzaku. We generated a large number of light curves by sampling the mean and standard deviation in each time step of the original light curves for both \suzaku\ and \x\ ({\bf Fig. S4b} (right)).  For each of the new light curves we then generated a power spectrum in an identical manner to the observed data and fit model~3 (Table S1). From the normalisation and equivalent width of the Lorentzian component used to model the QPO, it is then trivial to obtain the fractional r.m.s.~variability and an estimate of the error in this~value.    
\end{figure}

\begin{figure}[t]
\label{fig5}
\includegraphics[width=8.cm]{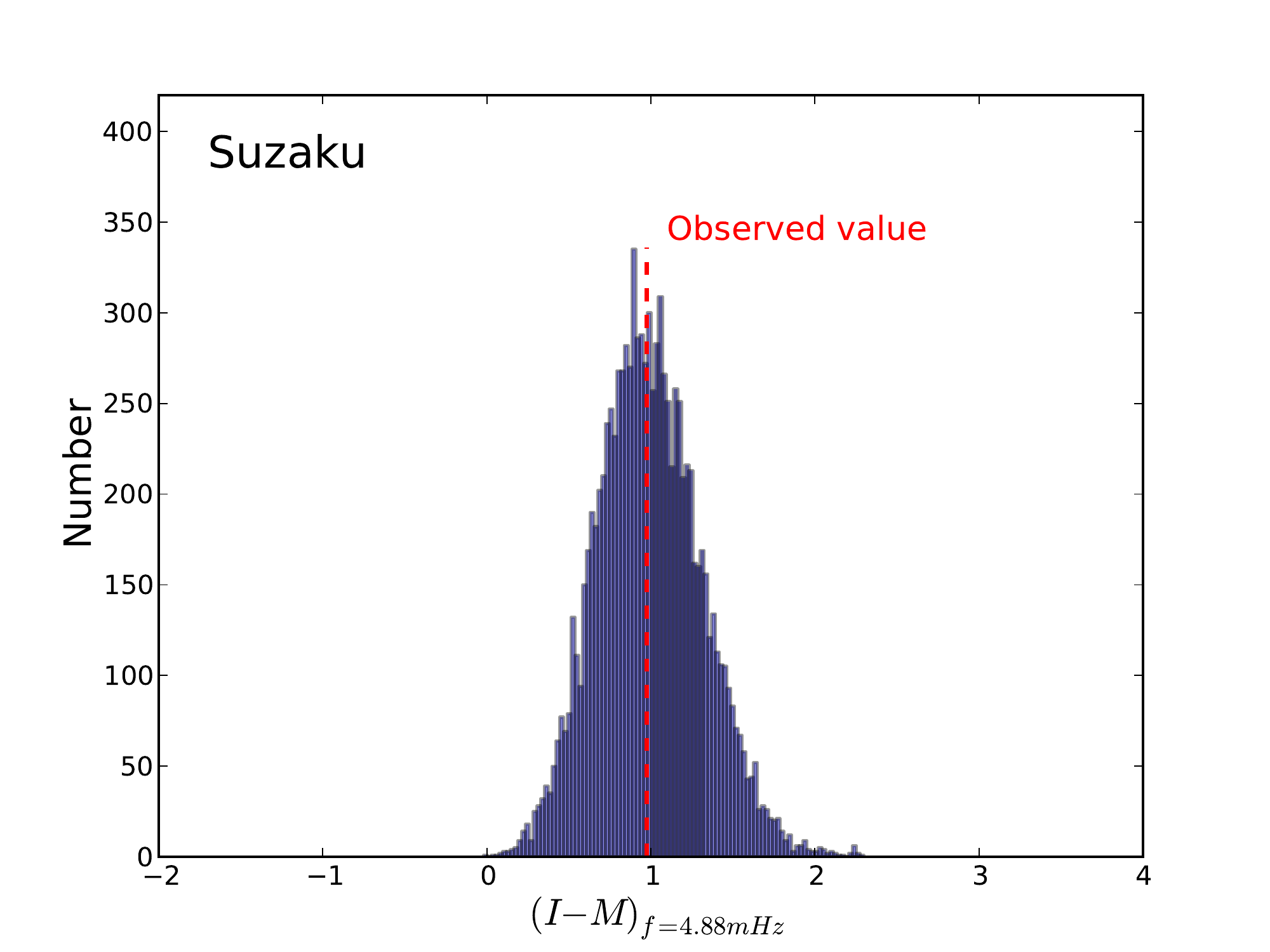}
\hspace*{0.cm}
\includegraphics[width=8.cm]{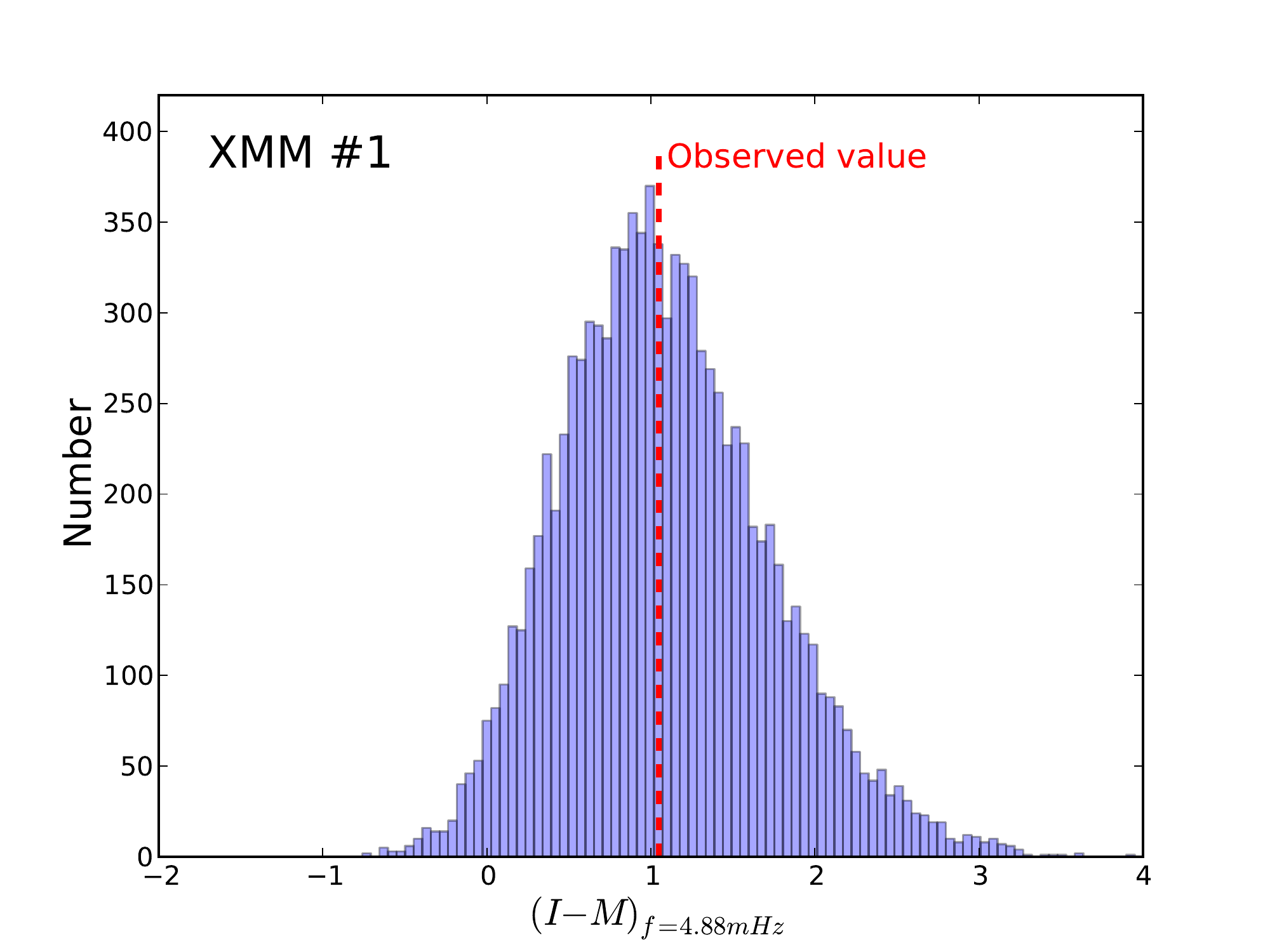}

\setlength{\baselineskip}{15.3pt} {\bf Fig. S5a} (left): Histogram of the power in the 4.88~mHz frequency bin in 10,000 bootstrap simulations for \suzaku.  The power (minus continuum level)  in the frequency bin with centroid frequency of 4.88~mHz in the 10,000 sampled spectra are shown as a blue histogram and it can be seen that it agrees well with the real observed value (red vertical dashed lines). This addresses the robustness of the signal in the QPO bin against possible statistical fluctuations.  {\bf Fig. S5b} (right): shows the same for \x.  
\end{figure}

Assuming that the signal is on top of a purely Poisson-noise time series of length $T$ and count-rate $I_x$, which is the product of $M$ individual power spectra, the limit on the r.m.s. variability ($r$)  found here, combined with Equation~2.17 of \cite{LewinvanParadijs1988SS}, results in a single-trial significance in the Gaussian limit $>3.8\sigma$ for \suzaku\ and $>2.2\sigma$ for \x.  However, in order to rigorously quantify the strength of this QPO outside of the Gaussian and Poissonian-noise limit as well as to account for all the frequencies probed in the various power spectra, we perform in the following section a full series of Monte Carlo simulations.

 Fig. S5 shows the distribution in the power\footnote{Throughout the rest of this work, use power, $I_f$ to refer to the signal strength in any frequency bin, $f$ when the power spectra is normalised such that their integral gives the squared r.m.s. fractional variability.} present in the QPO bin with centroid frequency of $4.88~$mHz after having the continuum-model level subtracted (i.e. $(I-M)_{f=4.88mHz}$). The continuum in each of the 10,000 simulated power spectra was found by fitting Model~2 (Table~S1) to each power spectrum.       The dashed red line shows that the observed value lies at the peak of the distribution in both observations, which confirms that the signal level in the QPO bin are not a statistical fluctuation in either observations.
 
\section{Monte Carlo Simulations} \label{MCsimulation}

The previous tests and quoted significances accounts for white noise variability; however, the presence of red noise is known to be a potential source of false features in the X-ray power spectra of X-ray binaries\cite{Vaughanqpo2010}. In order to fully investigate the significance of the aforementioned feature, we  implemented Monte Carlo techniques and compared the spread of 50,000 simulated power spectra to the observed power spectra in order to estimate the significance, in a similar manner to that outlined in Uttley, McHardy \& Papadakis\cite{UttleyMcHardy2002MNRAS}. 

We simulated 50,000 light curves using the method of Timmer \& Koenig\cite{TimmerKoenig1995}, for a broken power law continuum similar to that of Model~3 (Table~S1), having the low frequency index and  break frequency frozen at  their respective value.  The high frequency power law slope was incremented  between -2.4 and -3.2 in steps of 0.1 and the  simulated light curve was resampled and binned to match exactly the pattern of the observed light curve. Power spectra for all simulated light curves were produced in an identical manner to that described in the previous sections.

\begin{figure}[t]
\label{fig6}
\begin{center}
\hspace*{-0.5cm}
\includegraphics[width=8.cm]{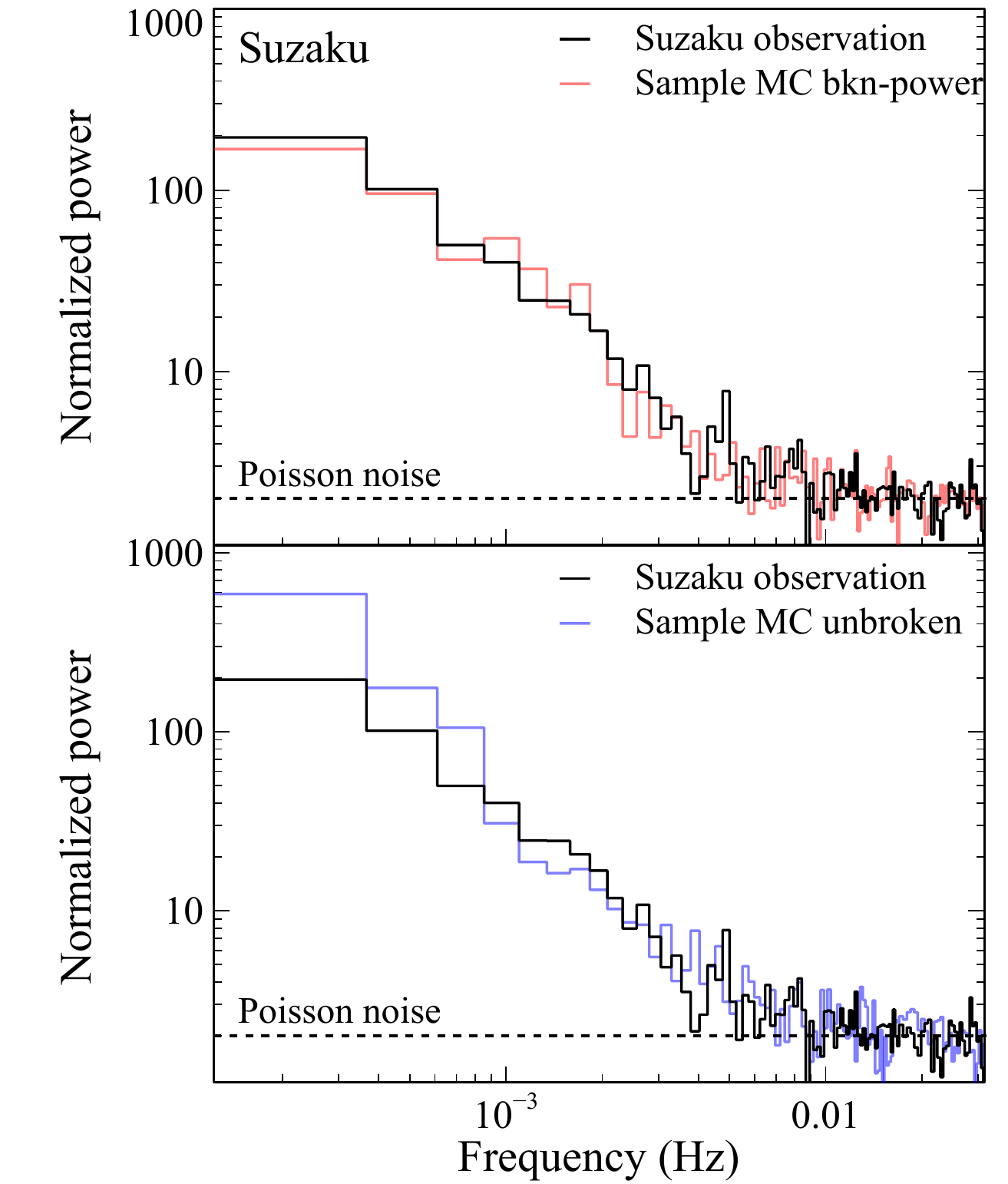}
\hspace*{-0.8cm}
\includegraphics[width=8.cm]{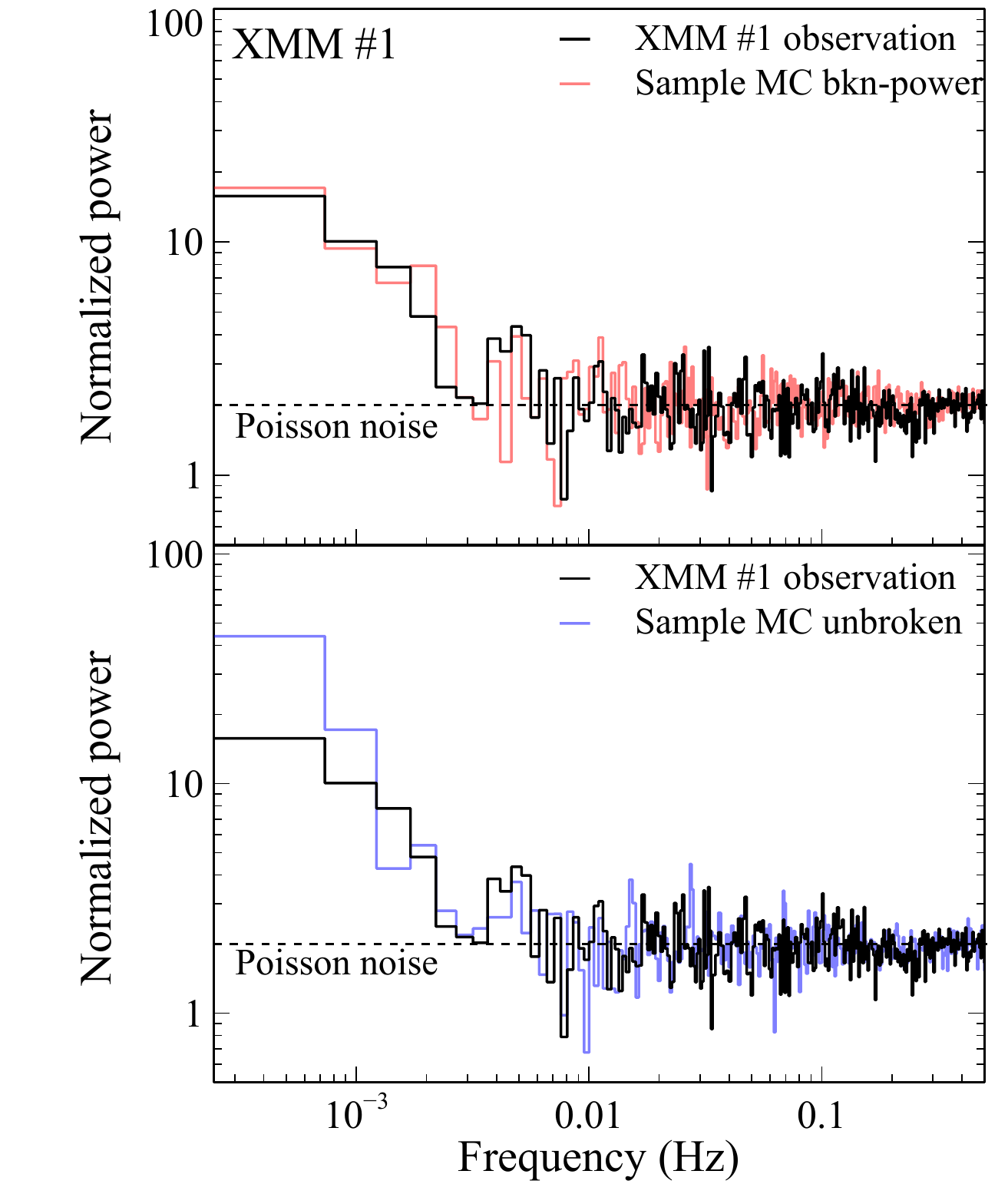}
\end{center}
\vspace*{-.5cm}
\setlength{\baselineskip}{15.3pt} {\bf Fig. S6} Sample simulated power spectrum from a broken (top) and unbroken (bottom) power law continuum. The broken power law provides a better description of both \textit{Suzaku} (left) and \textit{XMM}~\#1 (right)  data (shown in black; see also Table~S1). One of our statistical tests involves simulating 50,000 such spectra and comparing the results to the real data (see Figs.~S7--S9).
\end{figure}

\begin{figure}[h]
\label{fig7}
\begin{center}
\includegraphics[width=8.cm]{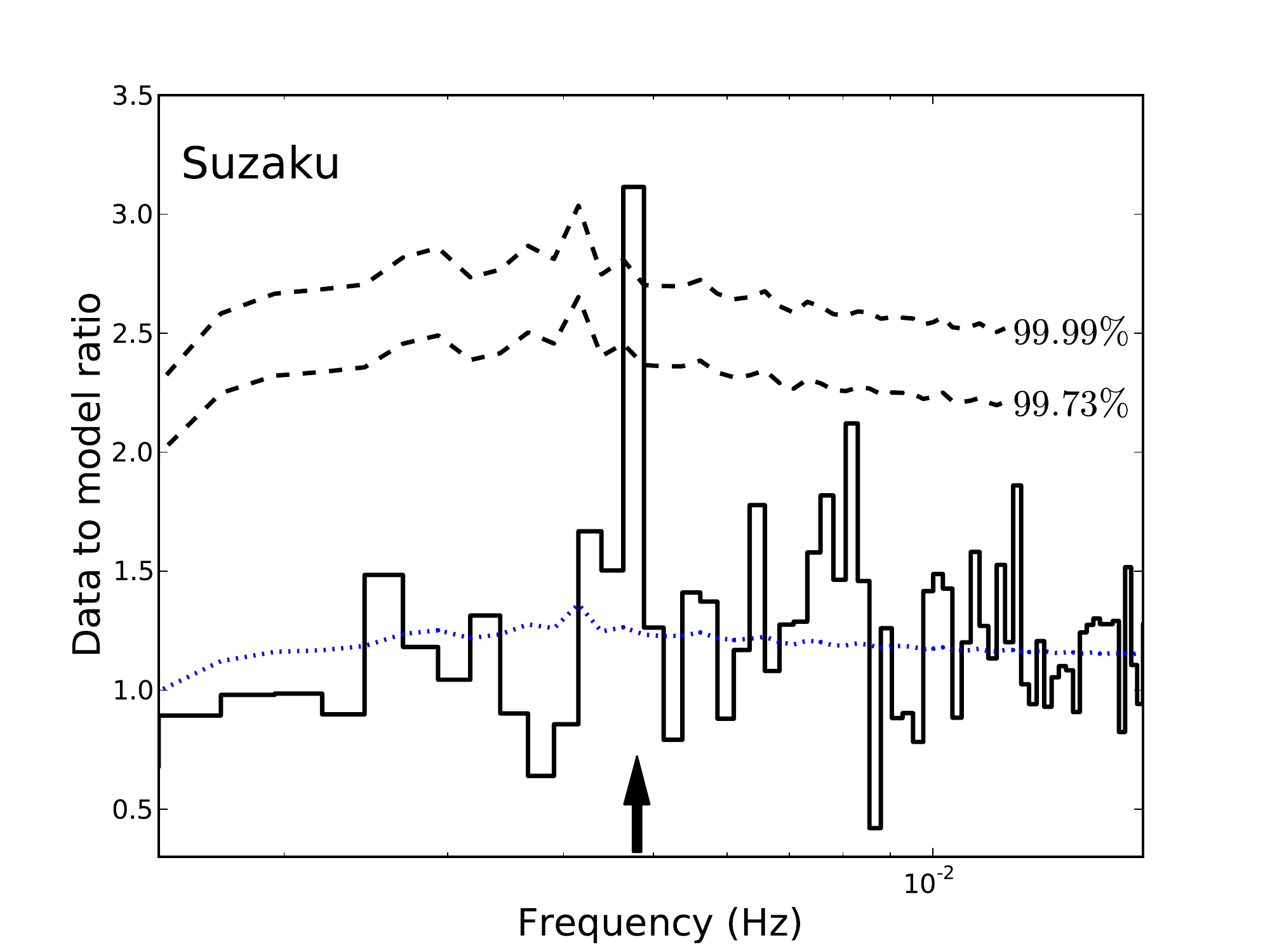}
\hspace*{-1.cm}
\includegraphics[width=8.cm]{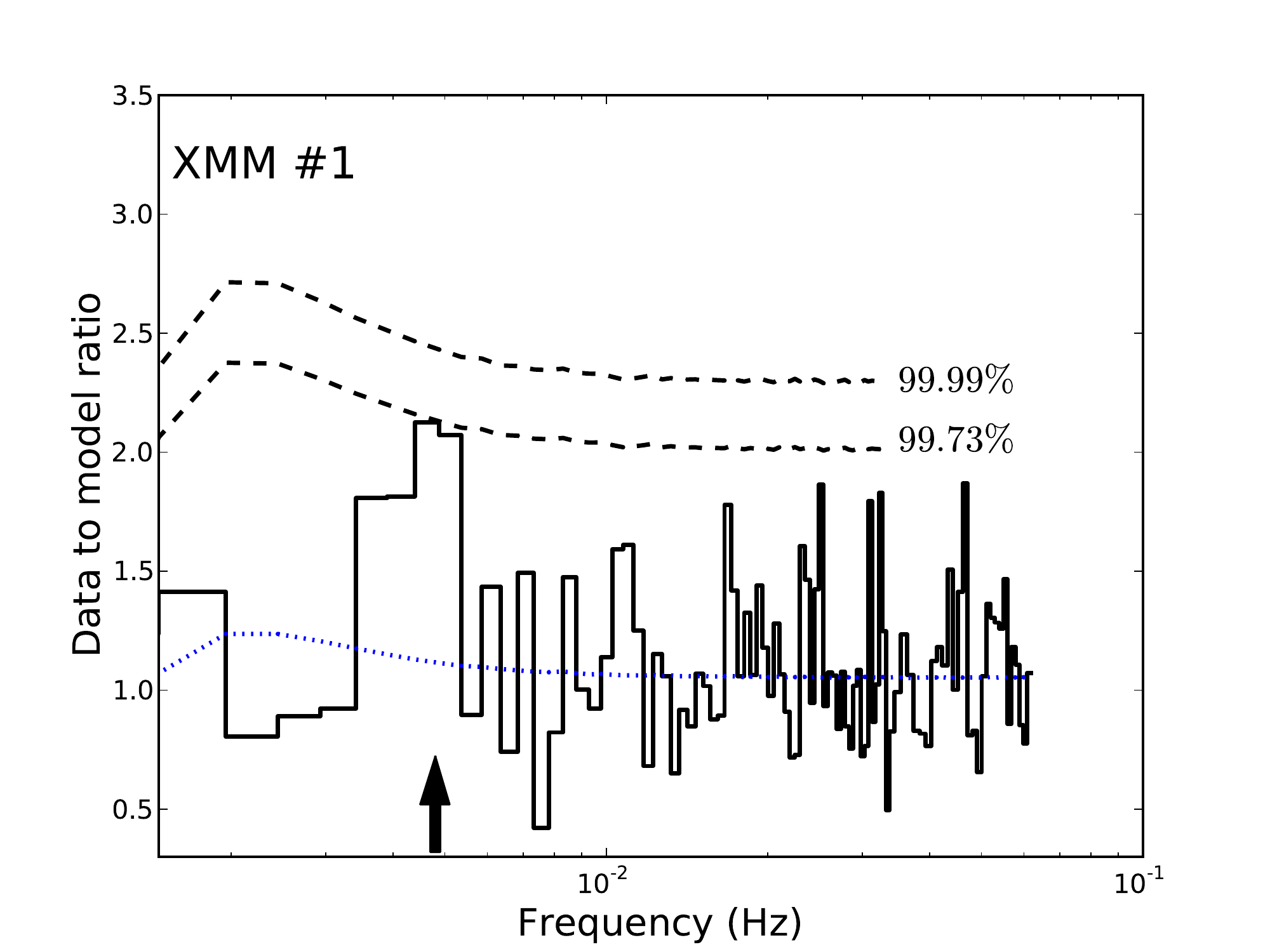}
\end{center}
\vspace*{-.5cm}
\setlength{\baselineskip}{15.3pt} {\bf Fig. S7a} (left): Data-to-model ratio for \textit{Suzaku}. {\bf Fig. S7b} (right) similarly for  \textit{XMM}~\#1. The ratio between the observed power spectra  and the continuum model (Model~2) are shown as the solid histogram. The dotted curve shows a similar ratio for the average of 50,000 Monte Carlo realisations to their respective continuum model and the dashed curves shows the $3\sigma$ (99.73\%) and 99.99\% confidence limits derived from the standard error obtained from the 50,000 realisations in each frequency channel. The arrow in both figures indicates the centroid frequency of the QPO as found in the \textit{Suzaku} data. It is clear that the \textit{XMM}~\#1 observation closely matches the frequency in the \textit{Suzaku}.
\end{figure}

\begin{figure}[h]
\label{fig8}
\includegraphics[width=8.5cm, height=6cm]{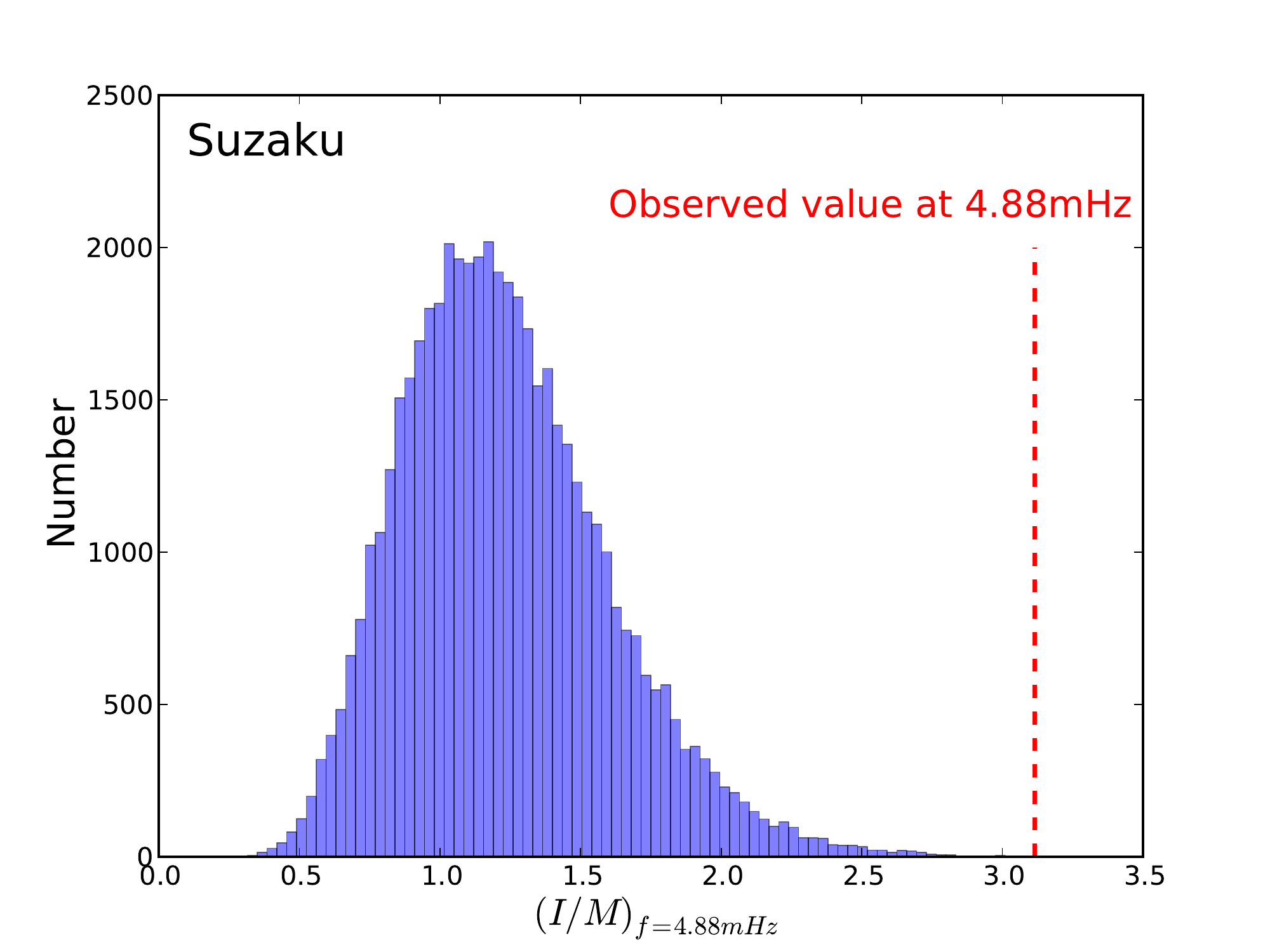}
\hspace*{-0.8cm}
\includegraphics[width=8.5cm, height=6cm]{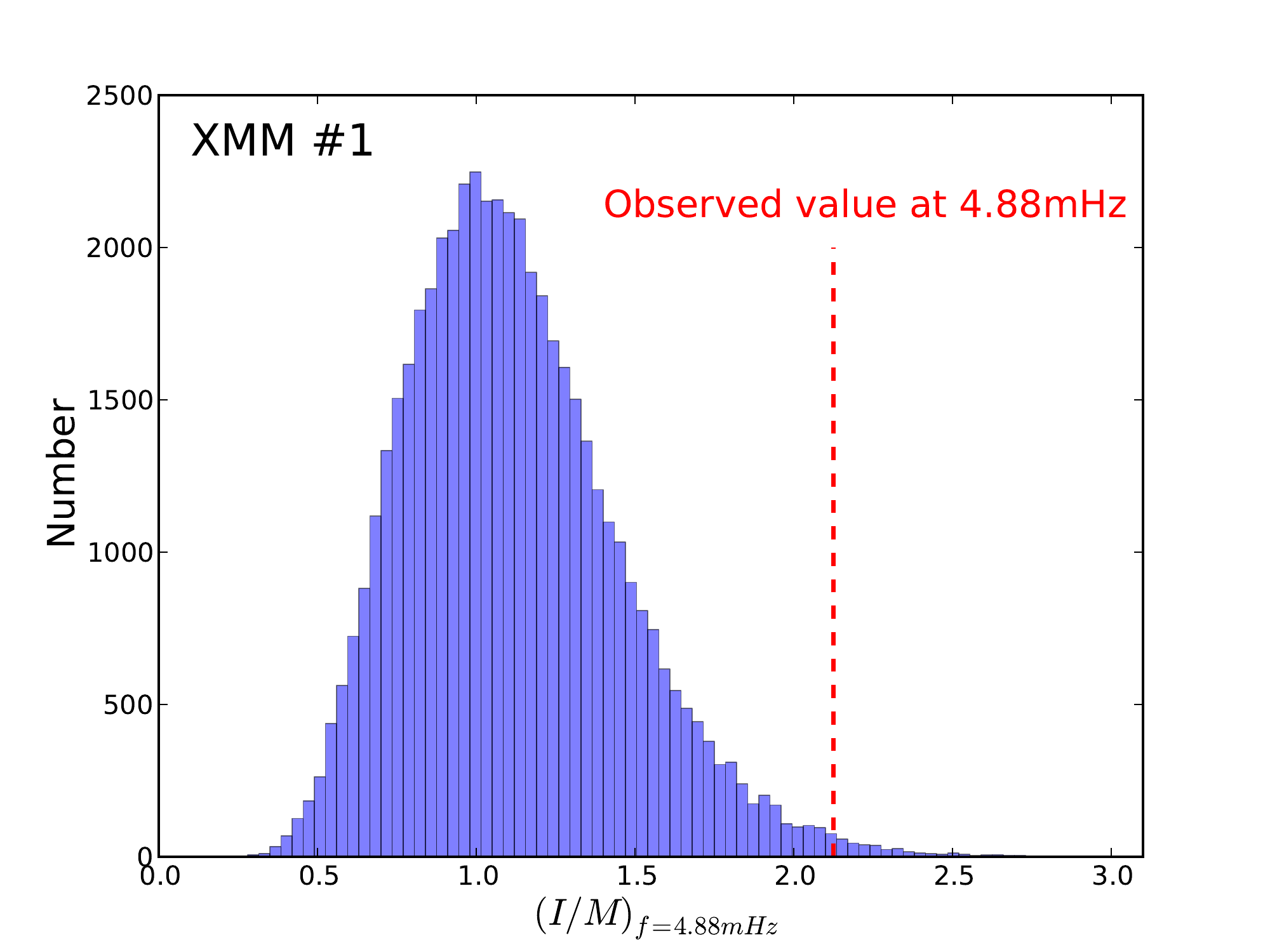}
\vspace*{-0.cm}
\includegraphics[width=8.5cm, height=6cm]{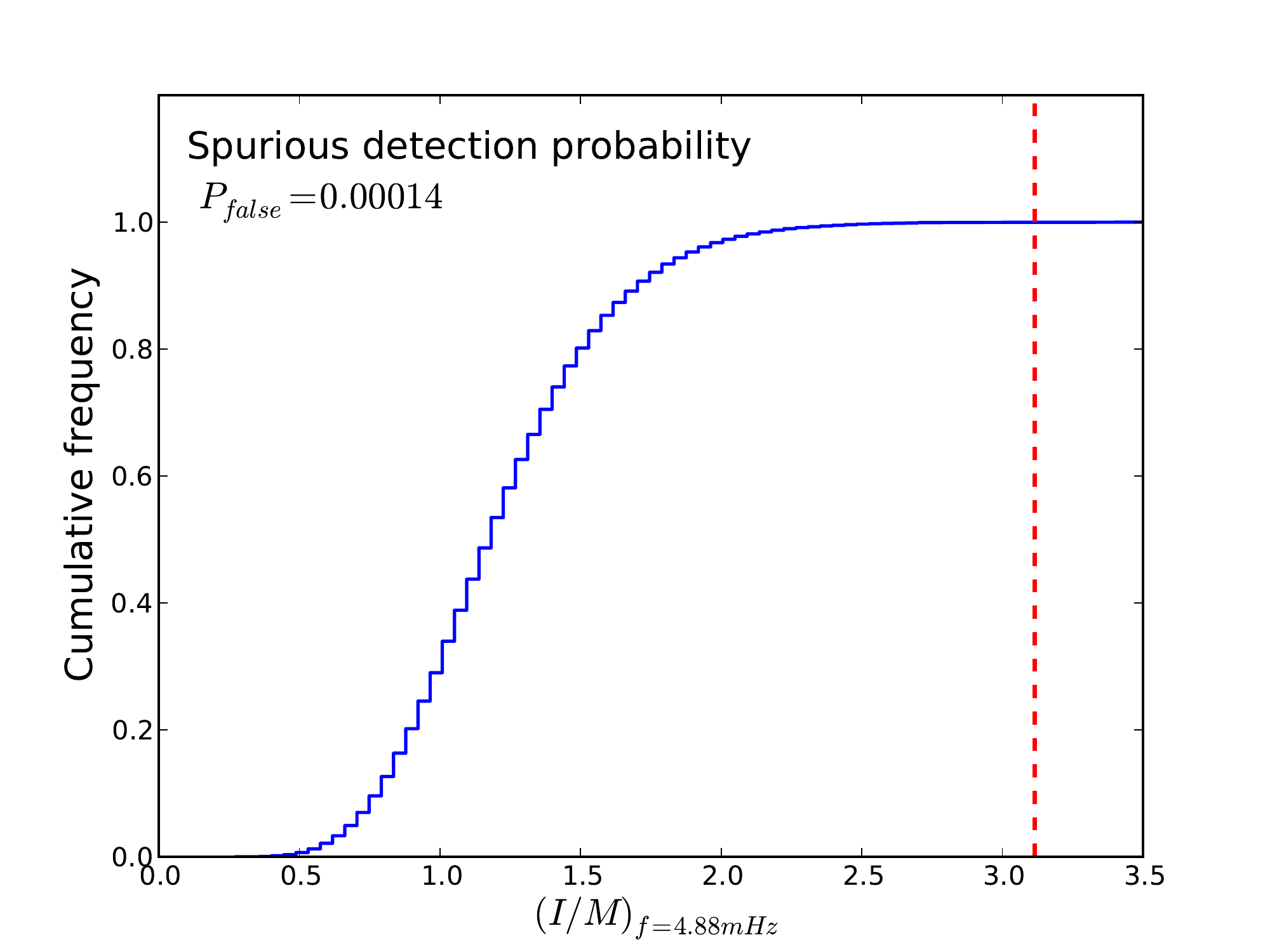}
\hspace*{-0.8cm}
\includegraphics[width=8.5cm, height=6cm]{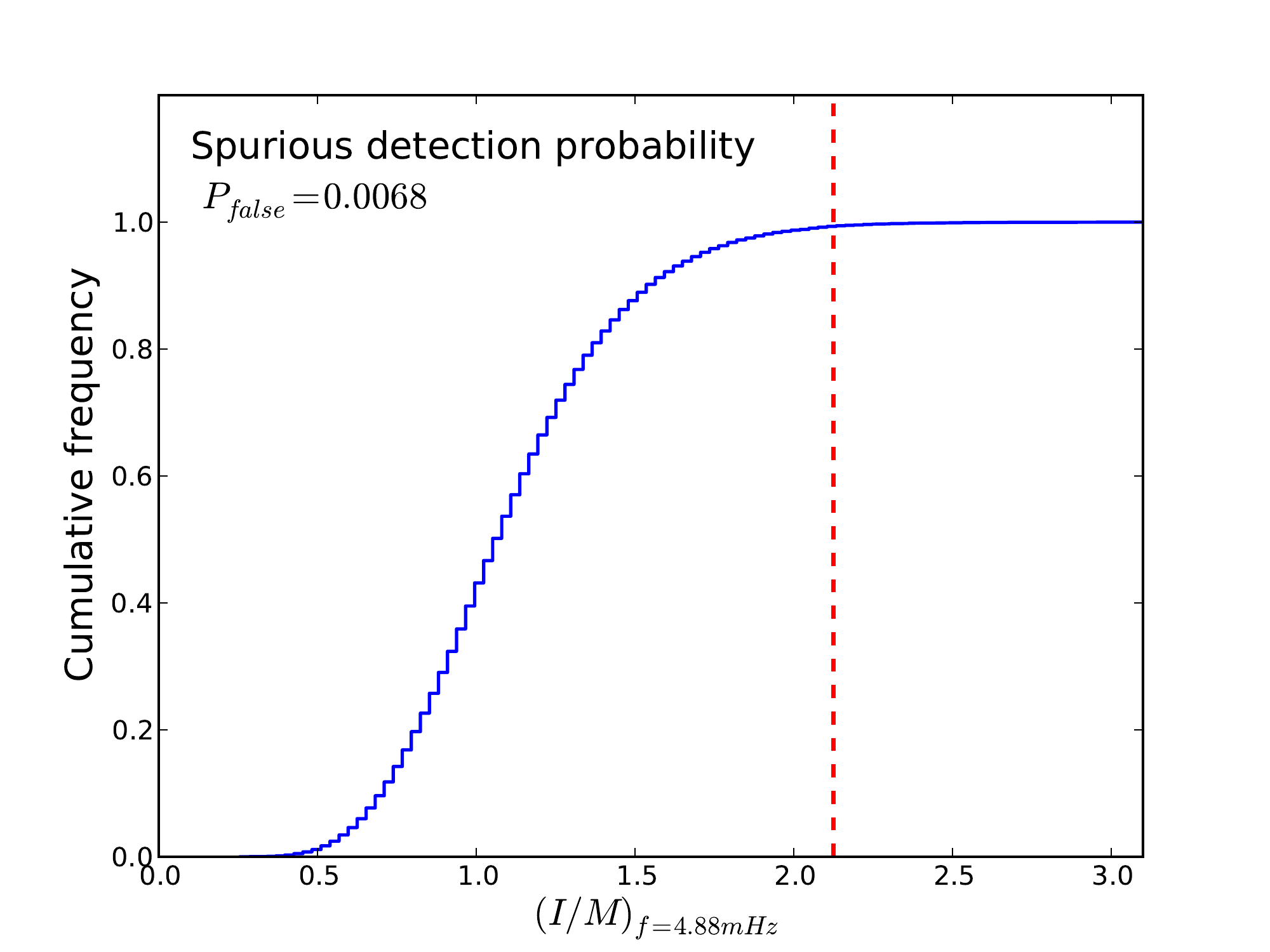}

\setlength{\baselineskip}{15.3pt} {\bf Fig. S8}: Histograms (top) and cumulative frequency (bottom) for the power in the QPO bin. The ratio of the power in the frequency bin with centroid frequency of 4.88~mHz  to the continuum model in the various Monte Carlo simulated spectra are shown as a histogram and cumulative frequency for \textit{Suzaku} (left) and \textit{XMM}~\#1 (right). For comparison, we show the real observed value as the red vertical dashed lines. The chance of  making a spurious detection (the false alarm probability) at the same power level as the \textit{Suzaku}~(\textit{XMM}~\#1) observed data at this  frequency is $1.4\times10^{-4}$~($6.8\times10^{-3}$) based on a total of 50,000 Monte Carlo simulated power spectra similar to that shown in Fig. S6 (top). Note however that this is  the probability for a single bin which, although appropriate for \textit{Suzaku}, is not the case for \textit{XMM}~\#1 where the QPO is clearly seen in at least two frequency bins (see Figs.~S7 and S9). 
\end{figure}

Fig. S6 (top) shows a characteristic simulated power spectrum (red) having continuum parameters similar to that of Model~3. Also shown is the real data (black) where we can  see that the simulated spectra indeed display similar noise level and characteristics as the original data. The bottom panels display a similar spectrum simulated for an unbroken power law (blue).  For each of the 50,000 Monte Carlo power spectrum simulated as described above,  we used Model~2 (Table~S1) to find the best fit to the red plus Poisson noise continuum. We show in Fig. S7 the data-to-model ratio for the observed data (solid histograms)  above the low frequency break as well as for the mean of all the simulations (blue dotted lines). We used Monte Carlo simulations to estimate 1~s.d errors in each frequency channel as exemplified  in Fig. S8 and show in Fig. S7 the $3\sigma$ (99.73\%) and 99.99\% confidence limits for the power as a function of frequency in the Monte Carlo simulation (dashed curves) .

\begin{figure}[t]
\label{fig9}
\begin{center}
\includegraphics[width=8.cm]{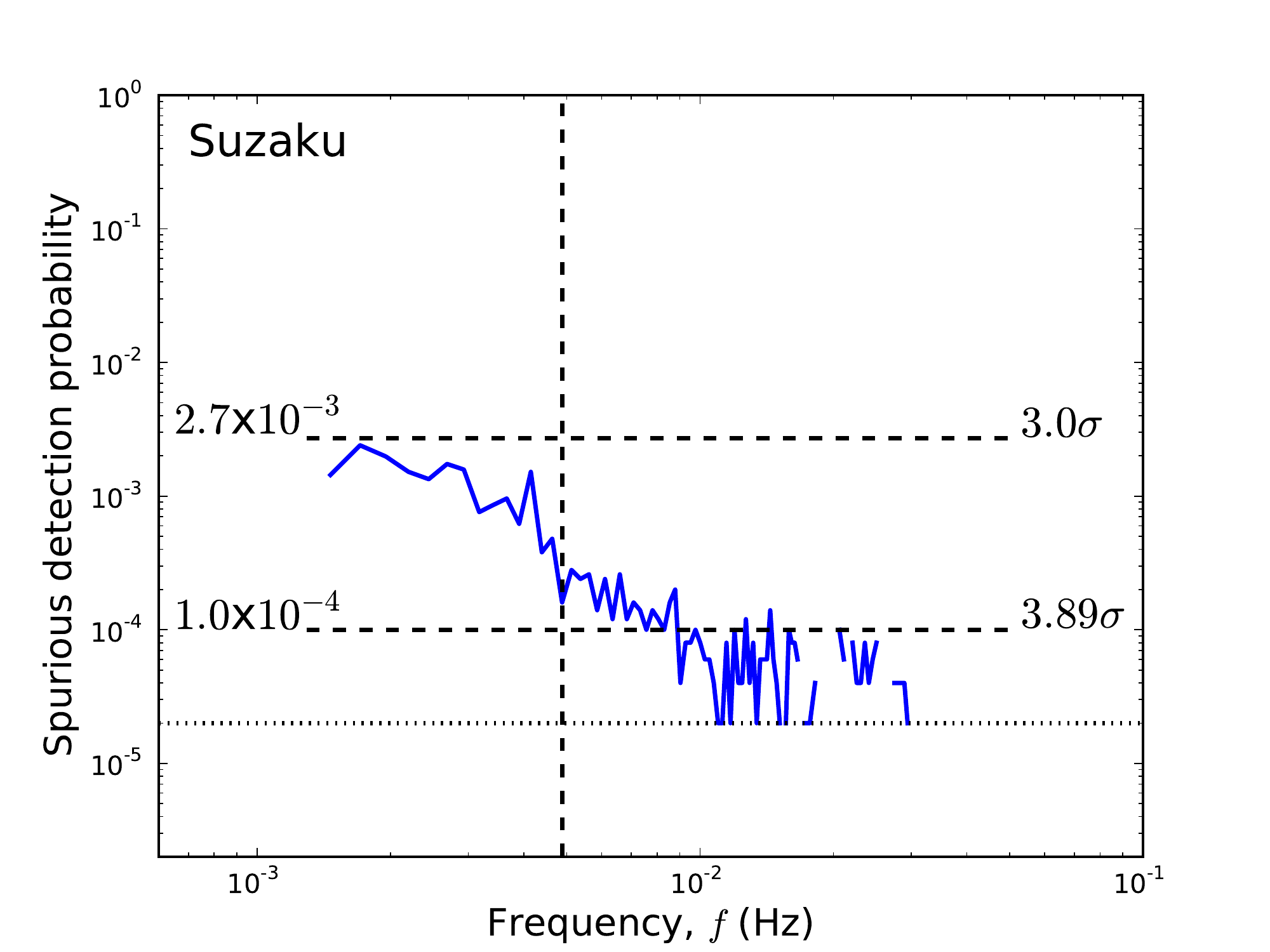}
\hspace*{-1.cm}
\includegraphics[width=8.cm]{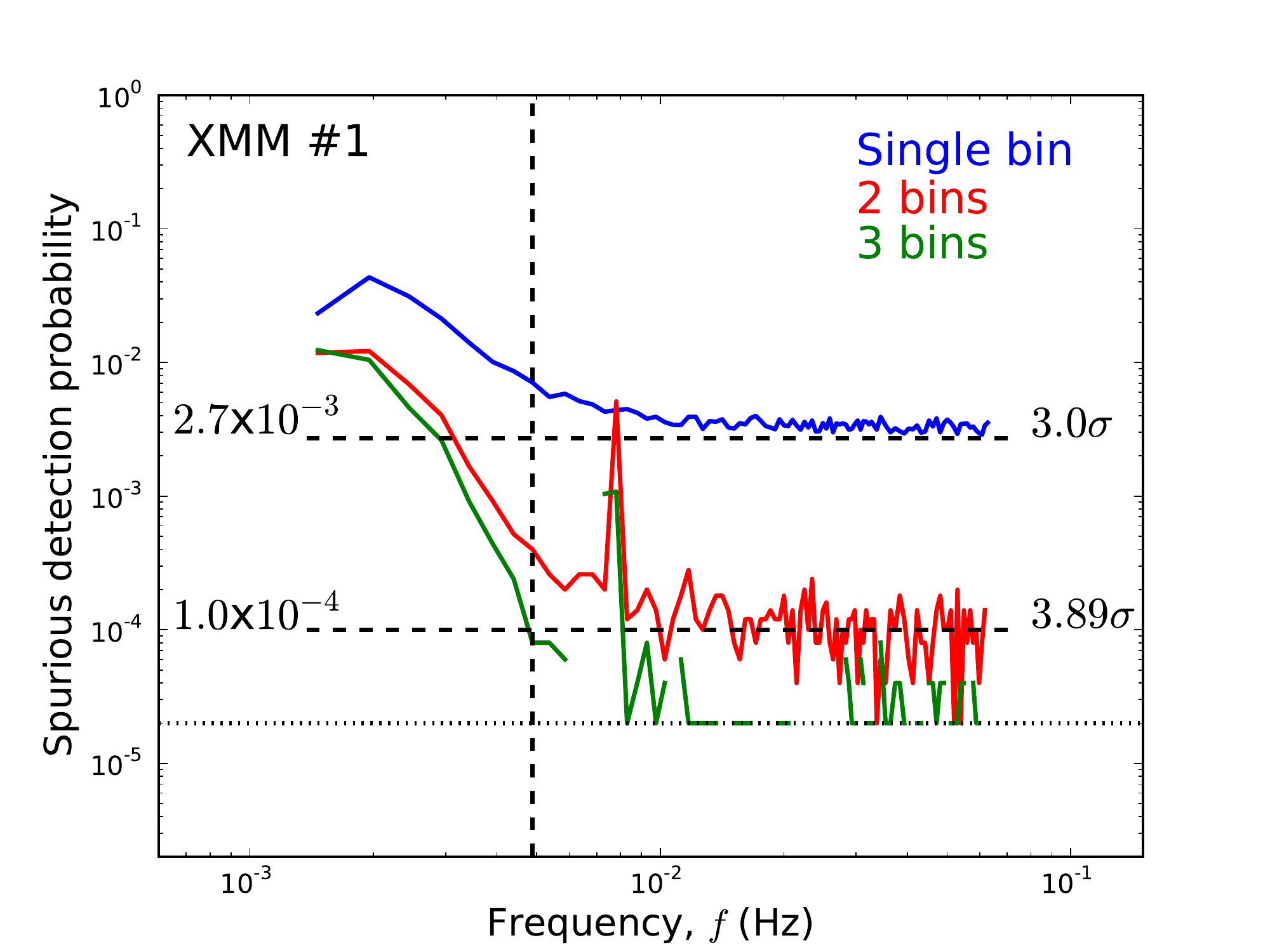}

\end{center}
\vspace*{-.0cm}
\setlength{\baselineskip}{15.3pt} {\bf Fig. S9a} (left): Spurious detection probability  for \textit{Suzaku}. For all frequency channels higher than the low frequency break (Fig. S6), we found the probability that the ratio between the power and continuum  at the specific frequency probed would be greater than the level found for the QPO in the real observation.  The vertical dashed line shows the centroid of the frequency bin containing the QPO and the horizontal dotted line shows the sensitive limit due to our limited number of simulations (50,000). The QPO is within a single bin in the \textit{Suzaku} observation (see Fig. S7) and we recover the value shown in Fig. S8 for \textit{Suzaku} at $\sim5$~mHz ($1.4\times10^{-4}$).  {\bf Fig.~S9b} (right): Shows a similar analyses for \x.  However, it is clear from Fig. S7 that the QPO in the \textit{XMM}~\#1 observation consists of multiple bins (at least 2) and indeed integrating over neighbouring bins we find that the chance of any two or three \textit{neighbouring} bins having the same integrated power as the QPO is significantly  low at $<5\times10^{-4}$. The significance levels are calculated assuming a Gaussian distribution from the percentage level shown on the left of the horizontal dashed lines. 
\end{figure}

As exemplified in Fig. S8, for each simulated power spectrum, we found the spread in the value $(I/M)_f$ and obtained the spurious-detection-probability, $P_{false}$, that this value would be at or greater than the value found in the real data at the frequency bin of the QPO. Fig. S8 shows the distribution for $(I/M)_{f=4.88mHz} $ with the observed value highlighted in red, and Fig. S9 summarises this result for all Fourier frequencies above the low frequency break. For 50,000 simulations, we found a power above the continuum at $\sim4.88$~mHz similar to that of the real data only 7 times for \suzaku. As hinted in Fig. S7, the QPO in the \textit{XMM}~\#1 observation, despite being at the same frequency as that found for \textit{Suzaku}, in fact occupies multiple frequency bins. This can be easily seen in  Fig. S7, where there are two neighbouring bins that reaches the 99.73\% curve. We show in Fig. S9 that the chances of any two (or 3) \textit{neighbouring} bins having a similar integrated power as the observed QPO bin is indeed very low at  $<5\times10^{-4}$.

\label{discussion}

So far we have calculated the probability of finding a QPO with the same power level as that found in the real data in a given frequency bin (i.e. Figures~S8 and S9). Since in general, when we first inspected the \textit{Suzaku} power spectrum, a QPO would have been reported at any plausible frequency we proceed by calculating the probability of chance occurrence $\textstyle \epsilon~=~1- \prod_{n=f_{min}}^{f_{max}} (1-P_{false|n})$, where $P_{false}$ is the false detection probability from Fig. S9,  of a detection at \textit{any} Fourier frequency greater than  $f_{min}$ which we have set to be the low frequency break for convenience. This product effectively accounts for our \textit{initial}  ``blind-search"  of the \textit{Suzaku} data and reduces the single observation significance to  $\epsilon= 2.31\times10^{-2} ~ (2.27\sigma)$. By itself, the \textit{Suzaku} observation would not constitute a discovery.

All our work so far however, has  only considered significance in individual observations. In both cases that we see the QPO, our extensive set of Monte Carlo simulations has shown that for any reasonable choice of binning and assumptions the single trial significance at the frequency of the QPO ($\sim5$~mHz) is greater than $3\sigma$. This is highlighted in Fig. S9.  We have shown here that the QPO exits in data based on two different telescopes which of course significantly reduces the likelihood that it is due to  instrumental systematic effects.  We have seen it in the two spectra with similar flux levels and spectral shape and at the same frequency. The latter is particularly important because when we embarked upon ``searching" for a QPO in the \textit{XMM}~\#1 data, we did not do so ``blindly".  Combining the blind-chance probability of a QPO in the \textit{Suzaku} data with the two-trials\footnote{ Using the 2 bins spurious detection probability in Fig. S9 and correcting for the fact that we effectively searched two frequency bins and not one.}  probability for a QPO at the \textit{same} frequency in the \textit{XMM}~\#1 data, $6.6\times10^{-4} ~( 3.41\sigma)$, we find the probability that these two detections arises from random noise alone is $1.52\times10^{-5}$, which assuming a Gaussianity equates to a detection at the $4.33\sigma$ level\footnote{Had we instead used the 3-neighbouring bins probability, accounting for the extra number of trial, we would have detection at the $4.43\sigma$ level of confidence. }. Thus the many statistical tests and simulations makes it clear that the QPO is robust and statistically significant.

\end{document}